\documentclass[aps,prl,reprint,times,floatfix,superscriptaddress]{revtex4-2}

\usepackage{amsmath}
\usepackage{amssymb}
\usepackage{bm}
\usepackage{graphicx,subfigure}
\usepackage{epsf,epsfig}
\usepackage{hyperref}
\usepackage{xcolor}
\usepackage{accents}

\begin{document}

\title{Mechano-chemical active feedback generates convergence extension in epithelial tissue}

\author{Aondoyima Ioratim-Uba}
\affiliation{School of Mathematics, University of Bristol - Bristol BS8 1UG, UK}
\author{Tanniemola B. Liverpool}
\affiliation{School of Mathematics, University of Bristol - Bristol BS8 1UG, UK}
\author{Silke Henkes}
\affiliation{School of Mathematics, University of Bristol - Bristol BS8 1UG, UK}
\affiliation{Lorentz Institute for Theoretical Physics, Leiden University - Leiden 2333 CA, The Netherlands}

\date{\today}

\begin{abstract}
Convergence extension, the simultaneous elongation of tissue along one axis while narrowing along a perpendicular axis, occurs during embryonic development. A fundamental process that contributes to shaping the organism, it happens in many different species and tissue types. 
 Here we present a minimal continuum model, that can be directly linked to the controlling microscopic biochemistry,  which shows spontaneous convergence extension. It is comprised of a 2D viscoelastic active material with a mechano-chemical active feedback mechanism coupled to a substrate via friction. Robust convergent extension behaviour emerges beyond a critical value of the activity parameter and is controlled by the boundary conditions and the coupling to the substrate. Oscillations and spatial patterns emerge in this model when internal dissipation dominates over friction, as well as in the active elastic limit.
 
\end{abstract}

\maketitle

\section{Introduction}

Convergent extension (CE) is a morphogenetic process that occurs during development. It is conserved across many different species, types of tissues and stages of development~\cite{huebner2018coming,keller2000mechanisms, bertet2004myosin, sutherland2020convergent}. During convergent extension, a region of sheet-like tissue (an epithelium) elongates in one direction (the long-axis) and contracts perpendicular to the long-axis. Convergent-extension plays a key role in a variety of developmental processes, such as primitive streak formation in chick embryos \cite{rozbicki2015myosin, saadaoui2020tensile} and drosophila germ band extension \cite{rauzi2008nature,rauzi2015embryo}. The formation of the primitive streak is an important part of gastrulation, the topological inversion process shared by nearly all multicellular animals and some plants \cite{hohn2015dynamics} that leads to cells taking up their correct positions within the embryo.

CE in epithelia is driven by cell intercalations~\cite{shindo2018models,keller2000mechanisms}, i.e. local cell rearrangements akin to the well-known topological T1 transitions of two-dimensional (passive) foams~\cite{durand2006relaxation, graner2008discrete}. Such T1s in passive systems relax stresses that build up from external driving and underlie the rheology of foams, which are typically yield stress materials \cite{weaire2008rheology}.  
However in epithelia, active T1 transitions \cite{rauzi2008nature,rauzi2015embryo, collinet2015local} can generate stresses locally even in the absence of external driving and can even develop local stresses that oppose external boundary forces. These are only possible due to motor-driven contractile stress generation, i.e. because the epithelial tissue is {\em active}. To obtain macroscopic strain against applied tension instead of a disordered response, the question then becomes how such events coordinate orientations with each other.

Until recently, the accepted answer has been a pre-existing morphogenetic gene expression pattern that bias local mechanical properties \cite{odell1980mechanical,lan2015biomechanical,shindo2018models}. However, the actomyosin fibres of the cytoskeleton are themselves susceptible to mechanical feedback \cite{koenderink2018architecture}, and in e.g. the chick embryo there is no evidence for pre-patterning. Therefore, more recent work has begun to include active feedback into models of one or several coupled junctions \cite{dierkes2014spontaneous,staddon2019mechanosensitive,cavanaugh2020adaptive}, and the response in tissues without T1s has also been investigated \cite{etournay2015interplay,noll2017active}.

\begin{figure}[t]
	\centering
	\includegraphics[width=0.5\textwidth]{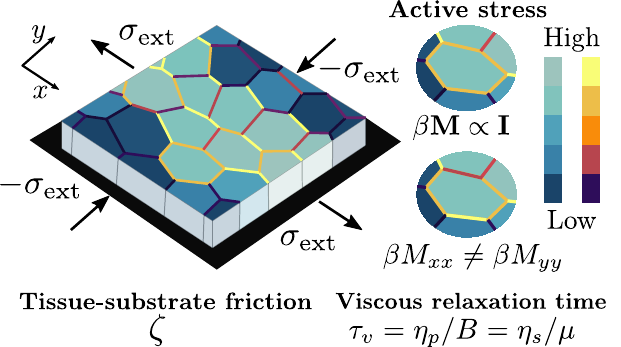}
	\caption{Schematic of a 2D sheet of tissue. The cell junctions are coloured purple/yellow/orange, to indicate the concentration of ActoMyosin: darker colour means less ActoMyosin. The active stress $\beta \bm{M}$ quantifies the concentration of and anisotropy of distribution of ActoMyosin in cells. The tissue is viscoelastic with a viscous relaxation time $\tau_v$ link to the bulk/shear viscosities $\eta_p$ and $\eta_s$, and the bulk and shear moduli $B$ and $\mu$ via $\tau_v = \eta_p/B = \eta_s/\mu$.}
	\label{fig:schematic}
\end{figure}

Then it becomes paramount to construct models of active tissue rheology with such feedback. While active models of cell sheets have a long tradition and include active gel theory \cite{kruse2005generic} and active nematic theories \cite{giomi2013defect}, the focus there has been on active instabilities and topological defect motion, including in in-vitro experiments \cite{saw2017topological}, but not on the response of the full tissue. Spatial patterns or direction of such feedback can be imposed \cite{streichan2018global, serra2021mechanochemical,chuai2023reconstruction,ibrahimi2022deforming} and the flow quantified, but so far a broader understanding of the emergent active relation between applied stress and strain rate, i.e. the tissue rheology, is missing.

\begin{figure}[t!]
	\centering
	\includegraphics[width = 0.5 \textwidth]{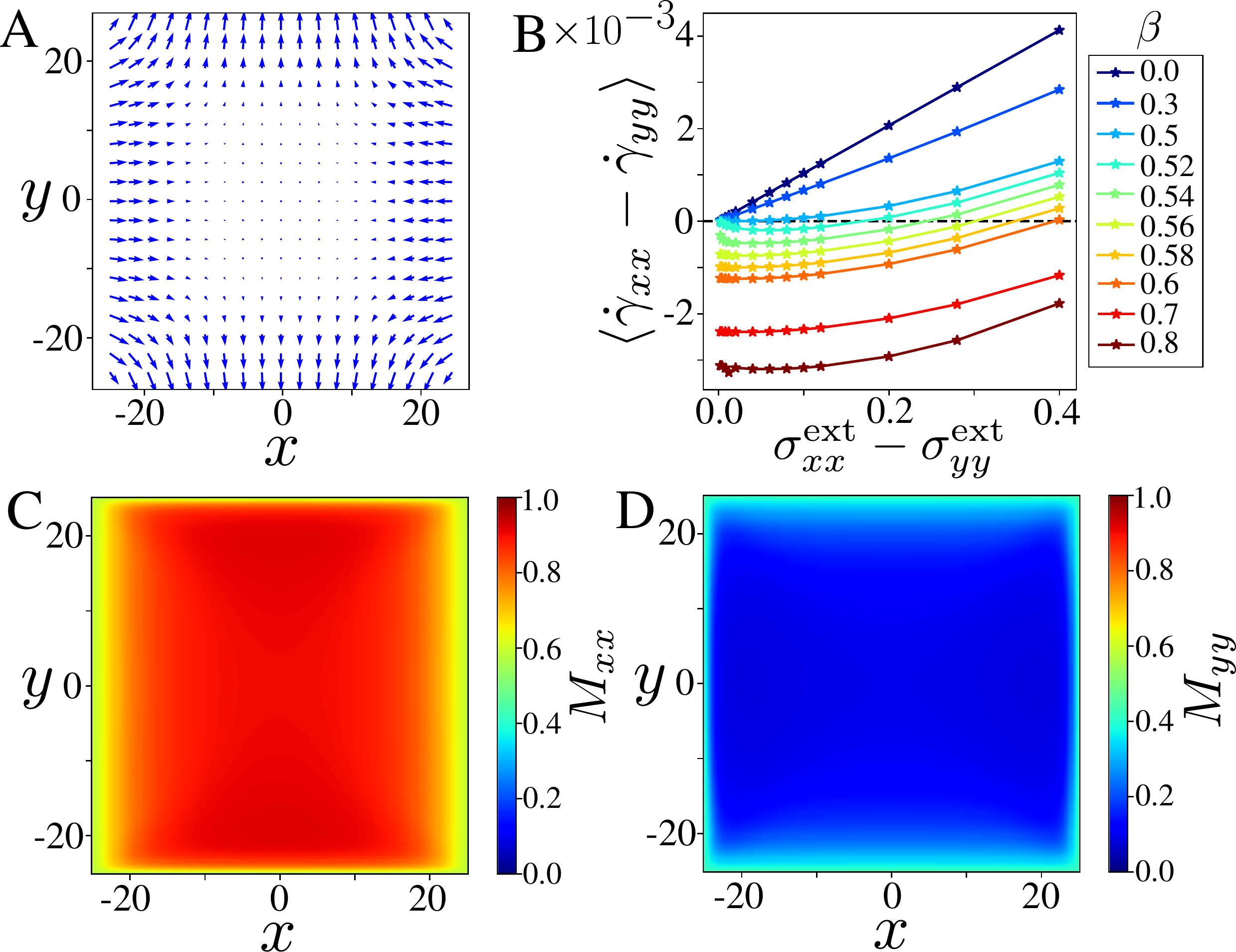}
	\caption {\textbf{A:} Steady state convergence extension velocity field at $\beta = 0.7$, $\sigma_{s} = 0.08$, and $\tau_m = \lambda = 20.0$. \textbf{B:} Pure shear strain as a function of pure shear stress at the boundary for various values of activity. \textbf{C:} $xx$ component and \textbf{D:} $yy$ component of the ActoMyosin tensor in the same convergence extension steady state as panel A.}
\label{fig:figure_2}
\end{figure}

In this letter, we present and analyse a continuum description of an epithelium with an active feedback mechanism where motor-driven contractile stress builds up, rather than relaxes,  in response to applied tension as in e.g. a catch-bond~\cite{sokurenko2008catch, veigel2003load}. This model is the continuum counterpart to a microscopic cell junction model~\cite{sknepnek2021generating} driven by stresses generated by molecular motors (myosin-II) and cytoskeletal filaments (F-actin) which is able to generate active T1s and limited C-E flow in a tissue patch. 
We formulate the model in terms of the distribution of ActoMyosin (myosin-II bound to F-actin) within cells, passive viscoelastic stress and the velocity via momentum balance with a substrate (see Fig. \ref{fig:schematic}). Numerically solving the continuum equations shows that above a critical activity C-E states appear, characterised by flow \emph{against} externally applied stress which acts like a mechanical signal (Fig. \ref{fig:figure_2}). We explain this using a steady-state approximation of the feedback dynamics, where high (low) boundary stresses select a high (low) ActoMyosin fixed point in the interior, and then build up a spatial gradient leading to flow. Separately, we find oscillating and patterned states in this model in the active elastic and low substrate friction limits. To linear order, we are able to show that our equations describe an active nematic coupled to a stress field above a critical activity, but an isotropic material below it.

{\bf Model.} We write down continuum equations for the epithelium as a 2D viscoelastic material that generates active stresses internally, and is coupled to a substrate via friction. The fundamental quantity of our framework is the anisotropic spatiotemporal distribution of ActoMyosin within cells, quantified by the 2nd rank tensor $\bm{M}({\bf{r}},t)$  (see Fig. \ref{fig:schematic}). It is symmetric but not traceless, i.e. for cells with isotropic ActoMyosin distributions, $\bm{M}$ is proportional to the identity $\bm{I}$. ActoMyosin, which in real tissues is distributed on the apical surface and along cell junctions, generates the stresses needed for cell-cell junction remodelling, which allows for active T1 transitions to occur. 
The other fields that characterise the material are local velocity ${\bf{v}}({\bf{r}}, t)$ and the local passive stress $\bm{\pi}({\bf{r}}, t)$. The total stress $\bm{\sigma}({\bf{r}}, t)$ in the material is the sum of the passive stress and an active stress proportional to $\bm{M}$ : $\bm{\sigma} = \bm{\pi} + \beta(\bm{M} - m_0\bm{I})$, where $\beta$ is the activity parameter and $m_0$ is the reference concentration for ActoMyosin. We use $m_0 = 1/2$ throughout. 

The dynamics of $\bm{M}({\bf{r}},t)$ is based on a model for a single contractile active junction that can remodel itself (see \cite{sknepnek2021generating} and SI eq. S1-4). In this model, the myosin dissociation constant decreases exponentially with tension, controlled by the susceptibility $k_0$. Changing the precise functional form does not affect behaviour qualitatively.
In addition, the ActoMyosin tensor is convected and rotated by the flow and we write
\begin{equation} \label{eq:myosin_equation}
\tau_m\accentset{\circ}{\bm{M}} = \bm{I} - (\bm{I} + e^{-k_0\bm{\sigma}})\cdot\bm{M} + D\nabla^2\bm{M}.
\end{equation}
The over circle represents the corotational derivative $\accentset{\circ}{\bm{A}} = \partial_{t}\bm{A} + {\bf{v}}\cdot\nabla \bm{A} + \bm{\omega}\cdot\bm{A} - \bm{A}\cdot\bm{\omega}$, where $\bm{\omega} = (1/2)(\nabla {\bf{v}} - (\nabla {\bf{v}})^T)$ is the vorticity tensor. We also include ActoMyosin diffusion with diffusion constant $D$.

The cell-cell junctions within the tissue are viscoelastic \cite{clement2017viscoelastic}, as is the tissue as a whole, with a time scale of stress relaxation \cite{khalilgharibi2019stress}. We use a convected compressible Maxwell model for the passive stress, superimposing separate Maxwell models for compression and shear deformations, 
\begin{equation} \label{eq:compressible_maxwell}
\bm{\pi} + \tau_v\accentset{\circ}{\bm{\pi}} = \frac{1}{2}\eta_p\text{Tr}(\dot{\bm{\gamma}})\bm{I}  + \eta_s\left(\dot{\bm{\gamma}} - \frac{1}{2}\text{Tr}(\dot{\bm{\gamma}})\bm{I}\right),
\end{equation}
where $\tau_v$ is the viscous relaxation time scale, $\eta_p$ is the bulk viscosity, $\eta_s$ is the shear viscosity, and the strain rate tensor is related to the velocity field via $\dot{\bm{\gamma}} =  (1/2)(\nabla {\bf{v}} + (\nabla {\bf{v}})^T)$. The bulk and shear moduli of the system are related to the relaxation time scale and the viscosities via $\tau_v = \eta_p /B = \eta_s/\mu$. The tissue is coupled to a substrate with friction coefficient $\zeta$ via momentum balance in the over-damped limit,
\begin{equation} \label{eq:momentum_balance}
\zeta{\bf{v}} = \nabla\cdot\bm{\sigma}.
\end{equation}

\begin{figure*}
	\centering
	\includegraphics[width = 0.7 \textwidth]{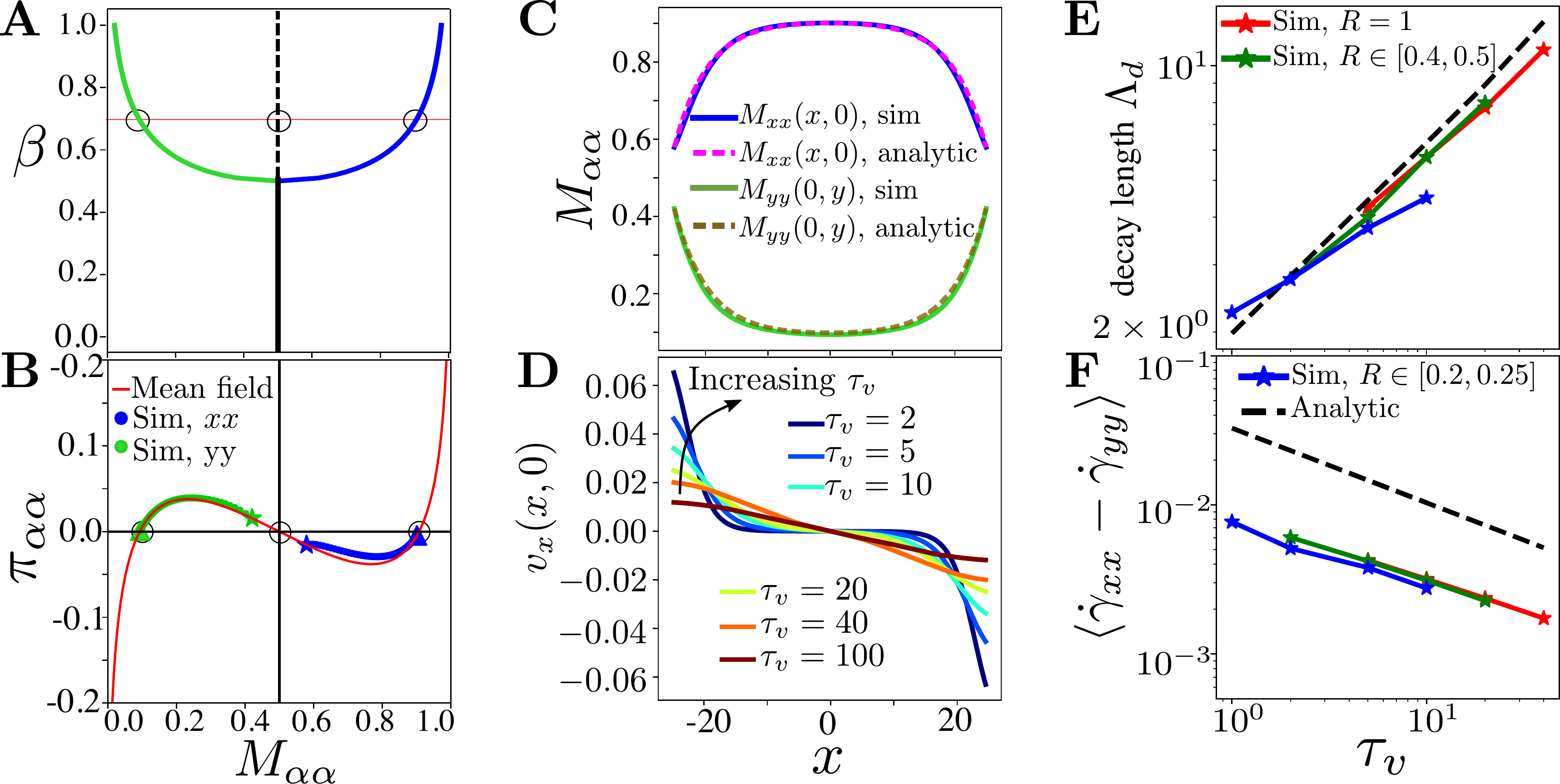}
	\caption{\textbf{A:} Pitchfork bifurcation of the mean field ActoMyosin concentration obtained via $\pi_{\alpha \alpha}(M_{\alpha \alpha}) = 0$ as function of activity. \textbf{B:} Mean field nullcline of $\dot{\bm{M}}$, $\pi_{\alpha \alpha}(M_{\alpha \alpha})$ (red) and corresponding simulation data, with blue dots for $\pi_{xx}(M_{xx})$ along $y=0$, green dots for $\pi_{yy}(M_{yy})$ along $x=0$. The central (boundary) points are marked with a triangle (stars), and the three  $\pi_{\alpha \alpha}(M_{\alpha \alpha}) = 0$ solutions with circles. \textbf{C:} $xx$ and $yy$ components of the simulated (solid) and mean-field (dashed) $\bm{M}$ tensor for the same parameters as fig. \ref{fig:figure_2}C. \textbf{D:} Simulated $v_x$ velocity profiles in the C-E state as a function of viscous time scale and for ratio $R=\tau_v/\tau_m\in [0.4-0.5]$, showing decay length. \textbf{E:} Decay length and \textbf{F:} pure strain rate in the C-E phase as a function of $\lambda$ together with mean field prediction (dashed).}  
\label{fig:figure_3}
\end{figure*}

{\bf Results. } We integrated the equations in time using the forwards Euler method, and approximated spatial derivatives using second order accurate finite difference on a square grid. 
The unit of time is set by the substrate elastic relaxation time scale $\tau_{el}=\zeta/B$, with both $\zeta=1$ and bulk modulus $B=1$, and we use a shear modulus $\mu=0.5$. The myosin feedback strength is set by $k_0=8$, with the exponential form of eq. (\ref{eq:myosin_equation}) limiting the components of $\bm{M}$ to the range $0-1$; then the active coupling $\beta$ sets the active stress scale.  
We use a system size of $L=50$ cell units with a linear grid spacing of $0.25$ and we fix $D = 1$. 

We simulate a patch inside a larger tissue by imposing a constant total stress $\bm{\sigma}_{\text{ext}}$ on the boundary while material flows through freely. Note that our model is neither incompressible nor density conserving as cells can reshape in the 3rd dimension, and also can divide or be extruded \cite{rozbicki2015myosin}. We complement this with the equilibrium value $\bm{M}=(1+e^{-k_0\bm \sigma^{\text{ext}}})^{-1}$ of the ActoMyosin tensor at the boundary and invert for the passive stress $\bm{\pi} = \bm{\sigma} -\beta(\bm{M} - m_0\bm{I})$.  To study C-E, we impose pure shear boundary conditions with simultaneous tension along $x$ and compression along $y$ as shown in Fig. \ref{fig:schematic}, i.e. $\sigma^{\text{ext}}_{xx} = -\sigma^{\text{ext}}_{yy}$ and $\sigma^{\text{ext}}_{xy} = 0$, resulting in a pure shear stress $\sigma_s \equiv \sigma^{\text{ext}}_{xx}-\sigma^{\text{ext}}_{yy}$. 

Figure \ref{fig:figure_2} summarises our findings. We measure tissue response using the spatially averaged pure shear strain rate $\dot{\gamma}_s \equiv \langle \dot{\gamma}_{xx} - \dot{\gamma}_{yy}\rangle$ of the steady state. We can see in Fig. \ref{fig:figure_2}B that at zero activity, $\dot{\gamma}_s =  \sigma_s/\eta_s$ i.e. the tissue indeed behaves as a viscous liquid. Below a threshold activity $\beta_c = 0.5$, the tissue continues to flow in the direction of applied stress, and the effective viscosity stays positive. However, ActoMyosin is built up by active tension feedback and eventually overwhelms pulling when $\beta > \beta_c$. Fig. \ref{fig:figure_2}C-D show how $\bm{M}_{xx}$ builds up in the tension direction, while $\bm{M}_{yy}$ symmetrically drops in the compression direction. Above $\beta_c$, the tissue then shows convergence extension (Fig. \ref{fig:figure_2}A) with the axis of elongation along the direction of compression, i.e. the tissue flows \emph{against} the applied force. ActoMyosin gradients and hence tissue flow is strongest near the boundary and decays into the bulk.
The C-E rheological curves above $\beta_c$ in Fig. \ref{fig:figure_2}B are highly unusual: the tissue responds to $\sigma_s \rightarrow 0$ with a strongly symmetry broken C-E, showing that the applied stress acts like a mechanical signal. When $\sigma_s$ increases, the C-E response diminishes, until at a $\beta$-dependent value the tissue flow reverses into the direction of pulling. For pure stretch / compression boundary conditions, the tissue also contracts / expands above $\beta_c$ (see SI Fig. 1-4 for full spatial profiles).

{\bf Analysis.} We can understand the observed spontaneous CE by approximating the steady-state solutions of eq. (\ref{eq:myosin_equation}-\ref{eq:momentum_balance}). From setting $\accentset{\circ}{\bm{M}}=0$, we can derive the ActoMyosin nullcline equations
\begin{equation} \label{eq:nullclines}
\pi_{\alpha \alpha} = -\frac{1}{k_0}\log\left(M_{\alpha \alpha}^{-1} - 1\right) - \beta(M_{\alpha \alpha} - m_0),
\end{equation}
where $\alpha = x,y $ and the off-diagonal components decay to zero (Fig. \ref{fig:figure_3}B). The only fixed point of the viscoelastic passive stress is $\pi_{\alpha \alpha} = 0$, resulting in the transcendental equation $\pi_{\alpha \alpha}(M_{\alpha \alpha}) = 0$. Below $\beta_c = 0.5$, this equation has one stable solution, $M_{\alpha \alpha} = m_0$. Above the critical activity, there is a pitchfork bifurcation with two stable branches $M_{\alpha \alpha} = m^+ > m_0$ and $M_{\alpha \alpha} = m^- < m_0$, while the $M_{\alpha \alpha} = m_0$ branch becomes unstable (Fig. \ref{fig:figure_3}A). 

During C-E, the equal and opposite imposed boundary stresses select a pair of points (stars) on the nullcline that break symmetry, and at the center of the tissue, we find $M_{xx} = m^+$ and $M_{yy} = m^-$. The boundary conditions determine the branches in the sense that if we reverse tension and compression directions, we have $M_{xx} = m^-$ and $M_{yy} = m^+$ instead. Convergence extension flows are generated by the spatial gradients in stress between boundary and centre via eq. \eqref{eq:momentum_balance}. We empirically observe that the values of these stresses interpolate between boundary and centre points \emph{along} the $\pi_{\alpha \alpha}(M_{\alpha \alpha})$ nullcline (Fig. \ref{fig:figure_3}B).

We can derive an approximate solution for the C-E steady state by linearly expanding around the stable $\pi_{\alpha \alpha}(m^{\pm}) = 0$ fixed points and write $\pi_{xx}(M_{xx}) = \pi^{\prime}(m^{+}) \: m_x$, $\pi_{yy}(M_{yy}) = \pi^{\prime}(m^{-}) \:m_y$ where $ m_x=M_{xx}-m^{+}, m_y = M_{yy} - m^{-}$, and set the off diagonal components to zero. We thus eliminate the ActoMyosin equation and once we use \eqref{eq:momentum_balance} to write the strain rate, the stress equation to linear order in $m_x$ and $m_y$ becomes
\begin{align}
\pi^{\prime}(m^{+})(m_x + \tau_v \partial_t m_x) & = A_{+}\partial^2_xm_x + A_{-}\partial^2_ym_y, \nonumber \\
\pi^{\prime}(m^{-})(m_y + \tau_v\partial_t m_y) & = B_{+}\partial^2_ym_y + B_{-}\partial^2_xm_x, \label{eq:linear_passive}\\
0 & = \partial_x\partial_y(C_+m_x + C_-m_y) \nonumber
\end{align}
where constants $A_{\pm}, B_{\pm}, C_{\pm}$ are given by SI eq. S9. If we work in the limit $t \gg \tau_v$, we can neglect the time derivative resulting in coupled PDEs in $x$ and $y$ for $m_x(x,y)$ and $m_y(x,y)$. The final solution takes the form of a hyperbolic cosine in $x$ and in $y$,
\begin{equation}
m_x = C_x \cosh \left[\!\frac{x}{\Lambda_+}\!\right] + C_y \cosh\left[\!\frac{y}{\Lambda_{\pm}}\!\right],
\end{equation}
with the full solution and derivation given in SI eq. S10-S12 and where the prefactors of each term are set by the boundary conditions.
The length scale $\Lambda_+ \sim \sqrt{(\eta_s + \eta_p)/\zeta}$, and we can derive the precise decay length (eq. SI S14) $\Lambda_d \sim \sqrt{\tau_v}$, independent of $\tau_m$ of the $\bm{M}_{\alpha\alpha}$ profiles.
Fig. \ref{fig:figure_3}C shows that the analytic approximation closely matches the numerical solution, and Fig. \ref{fig:figure_3}E shows that both the value of $\Lambda_d$ and the fact that it is $\tau_m$-independent are good prediction.
The same ratio of viscosity to substrate friction then determines the penetration length of the gradient and therefore the extent of C-E flow, as can be seen in the numerical velocity profiles in Fig. \ref{fig:figure_3}D. We can derive an analytical prediction for the C-E strain rate $\dot{\gamma}_s$, SI eq. S15, shown as a dashed line together with the numerics in Fig. \ref{fig:figure_3}F, showing the same scaling.

\begin{figure}
\centering
\includegraphics[width=\columnwidth]{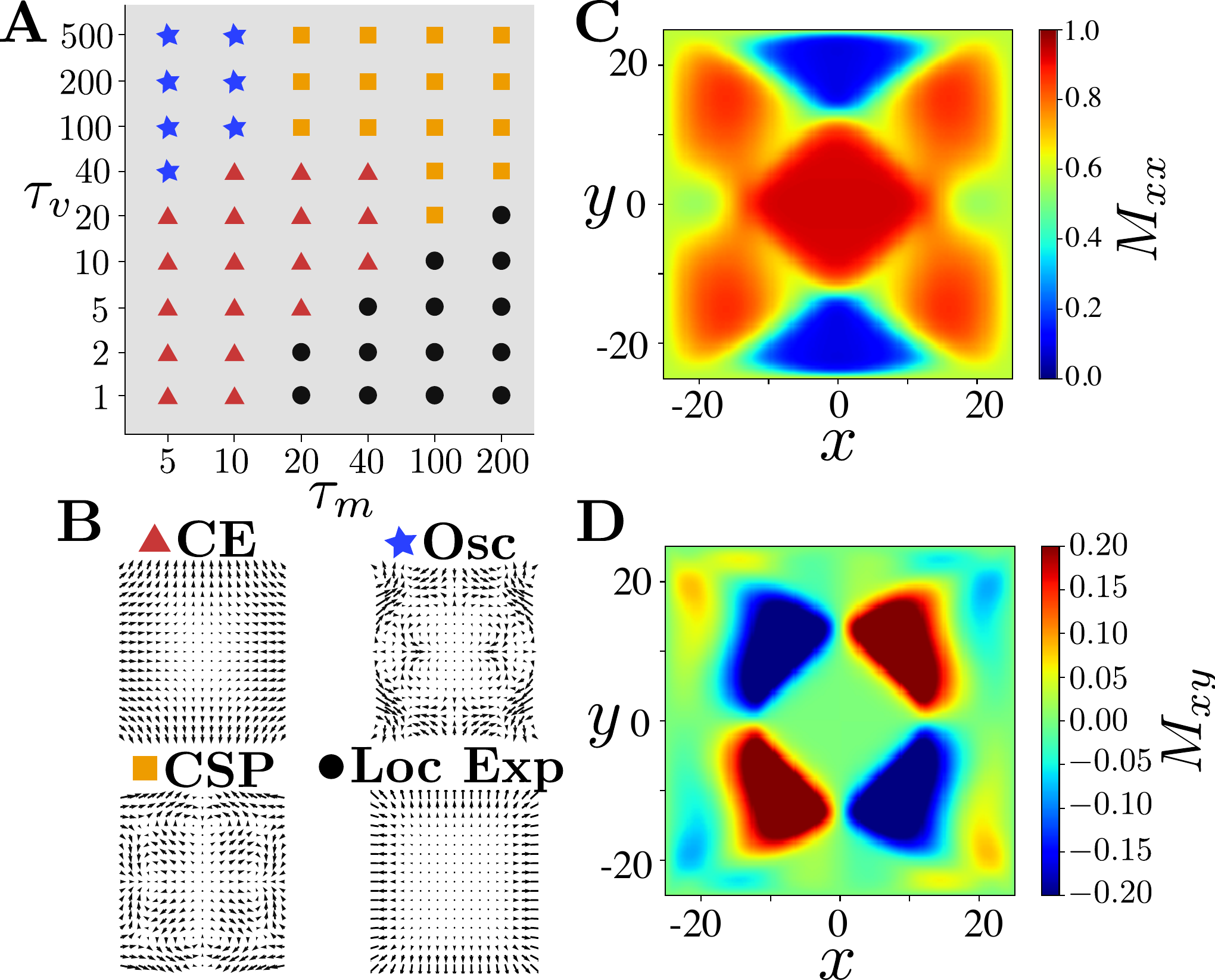}
\caption{Phases observed for other time scale ratios. \textbf{A:} Observed phases as a function of viscous relaxation time scale $\lambda$ and myosin time scale $\tau_m$, in units of elastic substrate relaxation time scale $\tau_{\text{el}}=\zeta/B=1$. \textbf{B:} Characteristic velocity fields of convergent-extension (C-E), instability, oscillating and localised expanding states. \textbf{C-D:} ActoMyosin wave pattern excited in the oscillating state for the  $M_{xx}$ (top) and off-diagonal $M_{xy}$ (bottom) components.}
\label{fig:phase_diagram}
\end{figure}

{\bf Broader context.} In phase diagram Figure \ref{fig:phase_diagram}A-B, we show that other solutions than C-E emerge in our model when we strongly increase either $\tau_v$ or $\tau_m$. For very small $\tau_v$ if $\tau_m$ is large, the solution as expected localises near the boundaries - however the state is expanding as $\bm{M}$ drops to the $m_-$ solution throughout (Fig. S7). If we instead take the limit $\tau_v \rightarrow \infty$ at small $\tau_m$, we observe pattern formation and regular oscillations in the system, including for the first time a significant $\bm{M}_{xy}$ component (Fig. \ref{fig:phase_diagram}C-D). This is the active \emph{elastic} limit where our system behaves as an elastic solid with moduli $B$ and $\mu$ coupled to the substrate with friction $\zeta$. It has recently been shown that active instabilities and odd (off-diagonal) responses are a characteristic feature of active elastic systems with feedback \cite{scheibner2020odd,brandenbourger2021limit} and that they also formally arise in viscoelastic systems \cite{banerjee2021active}. Here we show that they arise in a model for a biological tissue, raising the intriguing possibility that pattern formation in development could make use of such mechanisms.

In the limit of where $\tau_m,\tau_v \gg \tau_{el}$ (corresponding to the `wet' limit where substrate friction can be neglected), we observe a spatial destabilisation of the C-E pattern with slow dynamics that we have yet to fully explore (Fig. S6). 


Our governing equations bear some similarities to (and differences from)
the equations of active nematic systems. However unlike in those systems, we do not find the generic instabilities usually observed. We observe instead robust and steady CE flows which is clearly very useful for biological functionality and control. While the friction with the substrate and the viscoelasticity act as stabilisers on short times and lengthscales, the key feature that keeps robust control  is the interplay between the non-zero stress boundary conditions and the mechano-chemical feedback of the ActoMyosin dynamics:

To explore this nice feature of the model and to compare with the equations of classical nematodynamics, it is helpful to consider the dynamics of the traceless part of the ActoMyosin tensor,  $\bm{Q}$, i.e. 
$\bm{M} = \bm{Q} + \frac{1}{2}\text{Tr}(\bm{M})\bm{I}$.
By expanding the matrix exponential in equation \eqref{eq:myosin_equation} to linear order in $\bm{Q}$, one can show that (see SI eq. S16-S27)
\begin{equation}
\accentset{\circ}{\bm{Q}} = a\bm{Q} + b\tilde{\bm{\pi}} + c\widetilde{\bm{\pi}\cdot\bm{Q}} + d\widetilde{\bm{Q}\cdot\bm{\pi}} + D\nabla^2\bm{Q} + \mathcal{O}(\bm{Q}^2) ,
\end{equation}
where $\tilde{\bm{\pi}}$ is the traceless part of $\bm{\pi}$, $\widetilde{\bm{\pi}\cdot\bm{Q}}$ is the traceless part of $\bm{\pi}\cdot\bm{Q}$, $\widetilde{\bm{Q}\cdot\bm{\pi}}$ is the traceless part of $\bm{Q}\cdot\bm{\pi}$, $a = \frac{1}{2}\beta k_0 - 2$, $b = (k_0/2)(1 - (1/2)\beta k_0 \text{Tr}(\bm{Q}^2))$, $c = k_0(1 - \beta k_0/4)$, and $d = \beta k_0^2/4$. The first term shows that there is an isotropic to nematic transition at $\beta = 1/2$,  the critical activity derived from our theory. For $\beta < 1/2$, we have a stable isotropic material. For $\beta > 1/2$, we have a nematic, however the second term coming from the feedback from the passive stress at leading order resembles an applied field that will depend on the boundary conditions. This field will in general suppresses instabilities. The higher order terms will decorate this base state and can lead to a variety of interesting dynamical states, and consistent with the simulations. 
\\
In summary, here we have introduced a continuum model of developmental tissues where convergence-extension flows arise wholly from mechanical feedback. We find robust C-E flows where applied tension acts like an external field to determine the flow direction, based on breaking the symmetry of spontaneous ActoMyosin polarisation. C-E then arises from the active stress profile between a central fixed point solution and the imposed boundaries. Our model also shows pattern formation and spontaneous oscillations in the active elastic limit. 
\\

{\bf Acknowledgments.} TBL acknowledges support of BrisSynBio, a BBSRC/EPSRC Advanced Synthetic Biology Research Centre (grant number
BB/L01386X/1). SH acknowledges support of BBSRC
grant BB/N009150/2 and the University of Leiden. AIU is funded by an EPSRC DTP studentship.
The authors would like to thank Ilyas Djafer Ch\'erif, Rastko Sknepnek and Cornelis J. Weijer for in-depth discussions.

\vspace{-3mm}

\bibliography{active_sheet_paper}

\begin{thebibliography}{37}%
\makeatletter
\providecommand \@ifxundefined [1]{%
 \@ifx{#1\undefined}
}%
\providecommand \@ifnum [1]{%
 \ifnum #1\expandafter \@firstoftwo
 \else \expandafter \@secondoftwo
 \fi
}%
\providecommand \@ifx [1]{%
 \ifx #1\expandafter \@firstoftwo
 \else \expandafter \@secondoftwo
 \fi
}%
\providecommand \natexlab [1]{#1}%
\providecommand \enquote  [1]{``#1''}%
\providecommand \bibnamefont  [1]{#1}%
\providecommand \bibfnamefont [1]{#1}%
\providecommand \citenamefont [1]{#1}%
\providecommand \href@noop [0]{\@secondoftwo}%
\providecommand \href [0]{\begingroup \@sanitize@url \@href}%
\providecommand \@href[1]{\@@startlink{#1}\@@href}%
\providecommand \@@href[1]{\endgroup#1\@@endlink}%
\providecommand \@sanitize@url [0]{\catcode `\\12\catcode `\$12\catcode
  `\&12\catcode `\#12\catcode `\^12\catcode `\_12\catcode `\%12\relax}%
\providecommand \@@startlink[1]{}%
\providecommand \@@endlink[0]{}%
\providecommand \url  [0]{\begingroup\@sanitize@url \@url }%
\providecommand \@url [1]{\endgroup\@href {#1}{\urlprefix }}%
\providecommand \urlprefix  [0]{URL }%
\providecommand \Eprint [0]{\href }%
\providecommand \doibase [0]{https://doi.org/}%
\providecommand \selectlanguage [0]{\@gobble}%
\providecommand \bibinfo  [0]{\@secondoftwo}%
\providecommand \bibfield  [0]{\@secondoftwo}%
\providecommand \translation [1]{[#1]}%
\providecommand \BibitemOpen [0]{}%
\providecommand \bibitemStop [0]{}%
\providecommand \bibitemNoStop [0]{.\EOS\space}%
\providecommand \EOS [0]{\spacefactor3000\relax}%
\providecommand \BibitemShut  [1]{\csname bibitem#1\endcsname}%
\let\auto@bib@innerbib\@empty
\bibitem [{\citenamefont {Huebner}\ and\ \citenamefont
  {Wallingford}(2018)}]{huebner2018coming}%
  \BibitemOpen
  \bibfield  {author} {\bibinfo {author} {\bibfnamefont {R.~J.}\ \bibnamefont
  {Huebner}}\ and\ \bibinfo {author} {\bibfnamefont {J.~B.}\ \bibnamefont
  {Wallingford}},\ }\href@noop {} {\bibfield  {journal} {\bibinfo  {journal}
  {Developmental cell}\ }\textbf {\bibinfo {volume} {46}},\ \bibinfo {pages}
  {389} (\bibinfo {year} {2018})}\BibitemShut {NoStop}%
\bibitem [{\citenamefont {Keller}\ \emph {et~al.}(2000)\citenamefont {Keller},
  \citenamefont {Davidson}, \citenamefont {Edlund}, \citenamefont {Elul},
  \citenamefont {Ezin}, \citenamefont {Shook},\ and\ \citenamefont
  {Skoglund}}]{keller2000mechanisms}%
  \BibitemOpen
  \bibfield  {author} {\bibinfo {author} {\bibfnamefont {R.}~\bibnamefont
  {Keller}}, \bibinfo {author} {\bibfnamefont {L.}~\bibnamefont {Davidson}},
  \bibinfo {author} {\bibfnamefont {A.}~\bibnamefont {Edlund}}, \bibinfo
  {author} {\bibfnamefont {T.}~\bibnamefont {Elul}}, \bibinfo {author}
  {\bibfnamefont {M.}~\bibnamefont {Ezin}}, \bibinfo {author} {\bibfnamefont
  {D.}~\bibnamefont {Shook}},\ and\ \bibinfo {author} {\bibfnamefont
  {P.}~\bibnamefont {Skoglund}},\ }\href@noop {} {\bibfield  {journal}
  {\bibinfo  {journal} {Philosophical Transactions of the Royal Society of
  London. Series B: Biological Sciences}\ }\textbf {\bibinfo {volume} {355}},\
  \bibinfo {pages} {897} (\bibinfo {year} {2000})}\BibitemShut {NoStop}%
\bibitem [{\citenamefont {Bertet}\ \emph {et~al.}(2004)\citenamefont {Bertet},
  \citenamefont {Sulak},\ and\ \citenamefont {Lecuit}}]{bertet2004myosin}%
  \BibitemOpen
  \bibfield  {author} {\bibinfo {author} {\bibfnamefont {C.}~\bibnamefont
  {Bertet}}, \bibinfo {author} {\bibfnamefont {L.}~\bibnamefont {Sulak}},\ and\
  \bibinfo {author} {\bibfnamefont {T.}~\bibnamefont {Lecuit}},\ }\href@noop {}
  {\bibfield  {journal} {\bibinfo  {journal} {Nature}\ }\textbf {\bibinfo
  {volume} {429}},\ \bibinfo {pages} {667} (\bibinfo {year}
  {2004})}\BibitemShut {NoStop}%
\bibitem [{\citenamefont {Sutherland}\ \emph {et~al.}(2020)\citenamefont
  {Sutherland}, \citenamefont {Keller},\ and\ \citenamefont
  {Lesko}}]{sutherland2020convergent}%
  \BibitemOpen
  \bibfield  {author} {\bibinfo {author} {\bibfnamefont {A.}~\bibnamefont
  {Sutherland}}, \bibinfo {author} {\bibfnamefont {R.}~\bibnamefont {Keller}},\
  and\ \bibinfo {author} {\bibfnamefont {A.}~\bibnamefont {Lesko}},\ }in\
  \href@noop {} {\emph {\bibinfo {booktitle} {Seminars in cell \& developmental
  biology}}},\ Vol.\ \bibinfo {volume} {100}\ (\bibinfo {organization}
  {Elsevier},\ \bibinfo {year} {2020})\ pp.\ \bibinfo {pages}
  {199--211}\BibitemShut {NoStop}%
\bibitem [{\citenamefont {Rozbicki}\ \emph {et~al.}(2015)\citenamefont
  {Rozbicki}, \citenamefont {Chuai}, \citenamefont {Karjalainen}, \citenamefont
  {Song}, \citenamefont {Sang}, \citenamefont {Martin}, \citenamefont
  {Kn{\"o}lker}, \citenamefont {MacDonald},\ and\ \citenamefont
  {Weijer}}]{rozbicki2015myosin}%
  \BibitemOpen
  \bibfield  {author} {\bibinfo {author} {\bibfnamefont {E.}~\bibnamefont
  {Rozbicki}}, \bibinfo {author} {\bibfnamefont {M.}~\bibnamefont {Chuai}},
  \bibinfo {author} {\bibfnamefont {A.~I.}\ \bibnamefont {Karjalainen}},
  \bibinfo {author} {\bibfnamefont {F.}~\bibnamefont {Song}}, \bibinfo {author}
  {\bibfnamefont {H.~M.}\ \bibnamefont {Sang}}, \bibinfo {author}
  {\bibfnamefont {R.}~\bibnamefont {Martin}}, \bibinfo {author} {\bibfnamefont
  {H.-J.}\ \bibnamefont {Kn{\"o}lker}}, \bibinfo {author} {\bibfnamefont
  {M.~P.}\ \bibnamefont {MacDonald}},\ and\ \bibinfo {author} {\bibfnamefont
  {C.~J.}\ \bibnamefont {Weijer}},\ }\href@noop {} {\bibfield  {journal}
  {\bibinfo  {journal} {Nature cell biology}\ }\textbf {\bibinfo {volume}
  {17}},\ \bibinfo {pages} {397} (\bibinfo {year} {2015})}\BibitemShut
  {NoStop}%
\bibitem [{\citenamefont {Saadaoui}\ \emph {et~al.}(2020)\citenamefont
  {Saadaoui}, \citenamefont {Rocancourt}, \citenamefont {Roussel},
  \citenamefont {Corson},\ and\ \citenamefont {Gros}}]{saadaoui2020tensile}%
  \BibitemOpen
  \bibfield  {author} {\bibinfo {author} {\bibfnamefont {M.}~\bibnamefont
  {Saadaoui}}, \bibinfo {author} {\bibfnamefont {D.}~\bibnamefont
  {Rocancourt}}, \bibinfo {author} {\bibfnamefont {J.}~\bibnamefont {Roussel}},
  \bibinfo {author} {\bibfnamefont {F.}~\bibnamefont {Corson}},\ and\ \bibinfo
  {author} {\bibfnamefont {J.}~\bibnamefont {Gros}},\ }\href@noop {} {\bibfield
   {journal} {\bibinfo  {journal} {Science}\ }\textbf {\bibinfo {volume}
  {367}},\ \bibinfo {pages} {453} (\bibinfo {year} {2020})}\BibitemShut
  {NoStop}%
\bibitem [{\citenamefont {Rauzi}\ \emph {et~al.}(2008)\citenamefont {Rauzi},
  \citenamefont {Verant}, \citenamefont {Lecuit},\ and\ \citenamefont
  {Lenne}}]{rauzi2008nature}%
  \BibitemOpen
  \bibfield  {author} {\bibinfo {author} {\bibfnamefont {M.}~\bibnamefont
  {Rauzi}}, \bibinfo {author} {\bibfnamefont {P.}~\bibnamefont {Verant}},
  \bibinfo {author} {\bibfnamefont {T.}~\bibnamefont {Lecuit}},\ and\ \bibinfo
  {author} {\bibfnamefont {P.-F.}\ \bibnamefont {Lenne}},\ }\href@noop {}
  {\bibfield  {journal} {\bibinfo  {journal} {Nature cell biology}\ }\textbf
  {\bibinfo {volume} {10}},\ \bibinfo {pages} {1401} (\bibinfo {year}
  {2008})}\BibitemShut {NoStop}%
\bibitem [{\citenamefont {Rauzi}\ \emph {et~al.}(2015)\citenamefont {Rauzi},
  \citenamefont {Krzic}, \citenamefont {Saunders}, \citenamefont {Krajnc},
  \citenamefont {Ziherl}, \citenamefont {Hufnagel},\ and\ \citenamefont
  {Leptin}}]{rauzi2015embryo}%
  \BibitemOpen
  \bibfield  {author} {\bibinfo {author} {\bibfnamefont {M.}~\bibnamefont
  {Rauzi}}, \bibinfo {author} {\bibfnamefont {U.}~\bibnamefont {Krzic}},
  \bibinfo {author} {\bibfnamefont {T.~E.}\ \bibnamefont {Saunders}}, \bibinfo
  {author} {\bibfnamefont {M.}~\bibnamefont {Krajnc}}, \bibinfo {author}
  {\bibfnamefont {P.}~\bibnamefont {Ziherl}}, \bibinfo {author} {\bibfnamefont
  {L.}~\bibnamefont {Hufnagel}},\ and\ \bibinfo {author} {\bibfnamefont
  {M.}~\bibnamefont {Leptin}},\ }\href@noop {} {\bibfield  {journal} {\bibinfo
  {journal} {Nature communications}\ }\textbf {\bibinfo {volume} {6}},\
  \bibinfo {pages} {1} (\bibinfo {year} {2015})}\BibitemShut {NoStop}%
\bibitem [{\citenamefont {H{\"o}hn}\ \emph {et~al.}(2015)\citenamefont
  {H{\"o}hn}, \citenamefont {Honerkamp-Smith}, \citenamefont {Haas},
  \citenamefont {Trong},\ and\ \citenamefont {Goldstein}}]{hohn2015dynamics}%
  \BibitemOpen
  \bibfield  {author} {\bibinfo {author} {\bibfnamefont {S.}~\bibnamefont
  {H{\"o}hn}}, \bibinfo {author} {\bibfnamefont {A.~R.}\ \bibnamefont
  {Honerkamp-Smith}}, \bibinfo {author} {\bibfnamefont {P.~A.}\ \bibnamefont
  {Haas}}, \bibinfo {author} {\bibfnamefont {P.~K.}\ \bibnamefont {Trong}},\
  and\ \bibinfo {author} {\bibfnamefont {R.~E.}\ \bibnamefont {Goldstein}},\
  }\href@noop {} {\bibfield  {journal} {\bibinfo  {journal} {Physical review
  letters}\ }\textbf {\bibinfo {volume} {114}},\ \bibinfo {pages} {178101}
  (\bibinfo {year} {2015})}\BibitemShut {NoStop}%
\bibitem [{\citenamefont {Shindo}(2018)}]{shindo2018models}%
  \BibitemOpen
  \bibfield  {author} {\bibinfo {author} {\bibfnamefont {A.}~\bibnamefont
  {Shindo}},\ }\href@noop {} {\bibfield  {journal} {\bibinfo  {journal} {Wiley
  Interdisciplinary Reviews: Developmental Biology}\ }\textbf {\bibinfo
  {volume} {7}},\ \bibinfo {pages} {e293} (\bibinfo {year} {2018})}\BibitemShut
  {NoStop}%
\bibitem [{\citenamefont {Durand}\ and\ \citenamefont
  {Stone}(2006)}]{durand2006relaxation}%
  \BibitemOpen
  \bibfield  {author} {\bibinfo {author} {\bibfnamefont {M.}~\bibnamefont
  {Durand}}\ and\ \bibinfo {author} {\bibfnamefont {H.~A.}\ \bibnamefont
  {Stone}},\ }\href@noop {} {\bibfield  {journal} {\bibinfo  {journal}
  {Physical review letters}\ }\textbf {\bibinfo {volume} {97}},\ \bibinfo
  {pages} {226101} (\bibinfo {year} {2006})}\BibitemShut {NoStop}%
\bibitem [{\citenamefont {Graner}\ \emph {et~al.}(2008)\citenamefont {Graner},
  \citenamefont {Dollet}, \citenamefont {Raufaste},\ and\ \citenamefont
  {Marmottant}}]{graner2008discrete}%
  \BibitemOpen
  \bibfield  {author} {\bibinfo {author} {\bibfnamefont {F.}~\bibnamefont
  {Graner}}, \bibinfo {author} {\bibfnamefont {B.}~\bibnamefont {Dollet}},
  \bibinfo {author} {\bibfnamefont {C.}~\bibnamefont {Raufaste}},\ and\
  \bibinfo {author} {\bibfnamefont {P.}~\bibnamefont {Marmottant}},\
  }\href@noop {} {\bibfield  {journal} {\bibinfo  {journal} {The European
  Physical Journal E}\ }\textbf {\bibinfo {volume} {25}},\ \bibinfo {pages}
  {349} (\bibinfo {year} {2008})}\BibitemShut {NoStop}%
\bibitem [{\citenamefont {Weaire}(2008)}]{weaire2008rheology}%
  \BibitemOpen
  \bibfield  {author} {\bibinfo {author} {\bibfnamefont {D.}~\bibnamefont
  {Weaire}},\ }\href@noop {} {\bibfield  {journal} {\bibinfo  {journal}
  {Current Opinion in Colloid \& Interface Science}\ }\textbf {\bibinfo
  {volume} {13}},\ \bibinfo {pages} {171} (\bibinfo {year} {2008})}\BibitemShut
  {NoStop}%
\bibitem [{\citenamefont {Collinet}\ \emph {et~al.}(2015)\citenamefont
  {Collinet}, \citenamefont {Rauzi}, \citenamefont {Lenne},\ and\ \citenamefont
  {Lecuit}}]{collinet2015local}%
  \BibitemOpen
  \bibfield  {author} {\bibinfo {author} {\bibfnamefont {C.}~\bibnamefont
  {Collinet}}, \bibinfo {author} {\bibfnamefont {M.}~\bibnamefont {Rauzi}},
  \bibinfo {author} {\bibfnamefont {P.-F.}\ \bibnamefont {Lenne}},\ and\
  \bibinfo {author} {\bibfnamefont {T.}~\bibnamefont {Lecuit}},\ }\href@noop {}
  {\bibfield  {journal} {\bibinfo  {journal} {Nature cell biology}\ }\textbf
  {\bibinfo {volume} {17}},\ \bibinfo {pages} {1247} (\bibinfo {year}
  {2015})}\BibitemShut {NoStop}%
\bibitem [{\citenamefont {Odell}\ \emph {et~al.}(1980)\citenamefont {Odell},
  \citenamefont {Oster}, \citenamefont {Burnside},\ and\ \citenamefont
  {Alberch}}]{odell1980mechanical}%
  \BibitemOpen
  \bibfield  {author} {\bibinfo {author} {\bibfnamefont {G.}~\bibnamefont
  {Odell}}, \bibinfo {author} {\bibfnamefont {G.}~\bibnamefont {Oster}},
  \bibinfo {author} {\bibfnamefont {B.}~\bibnamefont {Burnside}},\ and\
  \bibinfo {author} {\bibfnamefont {P.}~\bibnamefont {Alberch}},\ }\href@noop
  {} {\bibfield  {journal} {\bibinfo  {journal} {Journal of mathematical
  biology}\ }\textbf {\bibinfo {volume} {9}},\ \bibinfo {pages} {291} (\bibinfo
  {year} {1980})}\BibitemShut {NoStop}%
\bibitem [{\citenamefont {Lan}\ \emph {et~al.}(2015)\citenamefont {Lan},
  \citenamefont {Wang}, \citenamefont {Fernandez-Gonzalez},\ and\ \citenamefont
  {Feng}}]{lan2015biomechanical}%
  \BibitemOpen
  \bibfield  {author} {\bibinfo {author} {\bibfnamefont {H.}~\bibnamefont
  {Lan}}, \bibinfo {author} {\bibfnamefont {Q.}~\bibnamefont {Wang}}, \bibinfo
  {author} {\bibfnamefont {R.}~\bibnamefont {Fernandez-Gonzalez}},\ and\
  \bibinfo {author} {\bibfnamefont {J.~J.}\ \bibnamefont {Feng}},\ }\href@noop
  {} {\bibfield  {journal} {\bibinfo  {journal} {Physical biology}\ }\textbf
  {\bibinfo {volume} {12}},\ \bibinfo {pages} {056011} (\bibinfo {year}
  {2015})}\BibitemShut {NoStop}%
\bibitem [{\citenamefont {Koenderink}\ and\ \citenamefont
  {Paluch}(2018)}]{koenderink2018architecture}%
  \BibitemOpen
  \bibfield  {author} {\bibinfo {author} {\bibfnamefont {G.~H.}\ \bibnamefont
  {Koenderink}}\ and\ \bibinfo {author} {\bibfnamefont {E.~K.}\ \bibnamefont
  {Paluch}},\ }\href@noop {} {\bibfield  {journal} {\bibinfo  {journal}
  {Current opinion in cell biology}\ }\textbf {\bibinfo {volume} {50}},\
  \bibinfo {pages} {79} (\bibinfo {year} {2018})}\BibitemShut {NoStop}%
\bibitem [{\citenamefont {Dierkes}\ \emph {et~al.}(2014)\citenamefont
  {Dierkes}, \citenamefont {Sumi}, \citenamefont {Solon},\ and\ \citenamefont
  {Salbreux}}]{dierkes2014spontaneous}%
  \BibitemOpen
  \bibfield  {author} {\bibinfo {author} {\bibfnamefont {K.}~\bibnamefont
  {Dierkes}}, \bibinfo {author} {\bibfnamefont {A.}~\bibnamefont {Sumi}},
  \bibinfo {author} {\bibfnamefont {J.}~\bibnamefont {Solon}},\ and\ \bibinfo
  {author} {\bibfnamefont {G.}~\bibnamefont {Salbreux}},\ }\href@noop {}
  {\bibfield  {journal} {\bibinfo  {journal} {Phys. Rev. Lett.}\ }\textbf
  {\bibinfo {volume} {113}},\ \bibinfo {pages} {148102} (\bibinfo {year}
  {2014})}\BibitemShut {NoStop}%
\bibitem [{\citenamefont {Staddon}\ \emph {et~al.}(2019)\citenamefont
  {Staddon}, \citenamefont {Cavanaugh}, \citenamefont {Munro}, \citenamefont
  {Gardel},\ and\ \citenamefont {Banerjee}}]{staddon2019mechanosensitive}%
  \BibitemOpen
  \bibfield  {author} {\bibinfo {author} {\bibfnamefont {M.~F.}\ \bibnamefont
  {Staddon}}, \bibinfo {author} {\bibfnamefont {K.~E.}\ \bibnamefont
  {Cavanaugh}}, \bibinfo {author} {\bibfnamefont {E.~M.}\ \bibnamefont
  {Munro}}, \bibinfo {author} {\bibfnamefont {M.~L.}\ \bibnamefont {Gardel}},\
  and\ \bibinfo {author} {\bibfnamefont {S.}~\bibnamefont {Banerjee}},\
  }\href@noop {} {\bibfield  {journal} {\bibinfo  {journal} {Biophysical
  Journal}\ }\textbf {\bibinfo {volume} {117}},\ \bibinfo {pages} {1739}
  (\bibinfo {year} {2019})}\BibitemShut {NoStop}%
\bibitem [{\citenamefont {Cavanaugh}\ \emph {et~al.}(2020)\citenamefont
  {Cavanaugh}, \citenamefont {Staddon}, \citenamefont {Banerjee},\ and\
  \citenamefont {Gardel}}]{cavanaugh2020adaptive}%
  \BibitemOpen
  \bibfield  {author} {\bibinfo {author} {\bibfnamefont {K.~E.}\ \bibnamefont
  {Cavanaugh}}, \bibinfo {author} {\bibfnamefont {M.~F.}\ \bibnamefont
  {Staddon}}, \bibinfo {author} {\bibfnamefont {S.}~\bibnamefont {Banerjee}},\
  and\ \bibinfo {author} {\bibfnamefont {M.~L.}\ \bibnamefont {Gardel}},\
  }\href@noop {} {\bibfield  {journal} {\bibinfo  {journal} {Current opinion in
  genetics \& development}\ }\textbf {\bibinfo {volume} {63}},\ \bibinfo
  {pages} {86} (\bibinfo {year} {2020})}\BibitemShut {NoStop}%
\bibitem [{\citenamefont {Etournay}\ \emph {et~al.}(2015)\citenamefont
  {Etournay}, \citenamefont {Popovi{\'c}}, \citenamefont {Merkel},
  \citenamefont {Nandi}, \citenamefont {Blasse}, \citenamefont {Aigouy},
  \citenamefont {Brandl}, \citenamefont {Myers}, \citenamefont {Salbreux},
  \citenamefont {J{\"u}licher} \emph {et~al.}}]{etournay2015interplay}%
  \BibitemOpen
  \bibfield  {author} {\bibinfo {author} {\bibfnamefont {R.}~\bibnamefont
  {Etournay}}, \bibinfo {author} {\bibfnamefont {M.}~\bibnamefont
  {Popovi{\'c}}}, \bibinfo {author} {\bibfnamefont {M.}~\bibnamefont {Merkel}},
  \bibinfo {author} {\bibfnamefont {A.}~\bibnamefont {Nandi}}, \bibinfo
  {author} {\bibfnamefont {C.}~\bibnamefont {Blasse}}, \bibinfo {author}
  {\bibfnamefont {B.}~\bibnamefont {Aigouy}}, \bibinfo {author} {\bibfnamefont
  {H.}~\bibnamefont {Brandl}}, \bibinfo {author} {\bibfnamefont
  {G.}~\bibnamefont {Myers}}, \bibinfo {author} {\bibfnamefont
  {G.}~\bibnamefont {Salbreux}}, \bibinfo {author} {\bibfnamefont
  {F.}~\bibnamefont {J{\"u}licher}}, \emph {et~al.},\ }\href@noop {} {\bibfield
   {journal} {\bibinfo  {journal} {Elife}\ }\textbf {\bibinfo {volume} {4}},\
  \bibinfo {pages} {e07090} (\bibinfo {year} {2015})}\BibitemShut {NoStop}%
\bibitem [{\citenamefont {Noll}\ \emph {et~al.}(2017)\citenamefont {Noll},
  \citenamefont {Mani}, \citenamefont {Heemskerk}, \citenamefont {Streichan},\
  and\ \citenamefont {Shraiman}}]{noll2017active}%
  \BibitemOpen
  \bibfield  {author} {\bibinfo {author} {\bibfnamefont {N.}~\bibnamefont
  {Noll}}, \bibinfo {author} {\bibfnamefont {M.}~\bibnamefont {Mani}}, \bibinfo
  {author} {\bibfnamefont {I.}~\bibnamefont {Heemskerk}}, \bibinfo {author}
  {\bibfnamefont {S.~J.}\ \bibnamefont {Streichan}},\ and\ \bibinfo {author}
  {\bibfnamefont {B.~I.}\ \bibnamefont {Shraiman}},\ }\href@noop {} {\bibfield
  {journal} {\bibinfo  {journal} {Nature physics}\ }\textbf {\bibinfo {volume}
  {13}},\ \bibinfo {pages} {1221} (\bibinfo {year} {2017})}\BibitemShut
  {NoStop}%
\bibitem [{\citenamefont {Kruse}\ \emph {et~al.}(2005)\citenamefont {Kruse},
  \citenamefont {Joanny}, \citenamefont {J{\"u}licher}, \citenamefont {Prost},\
  and\ \citenamefont {Sekimoto}}]{kruse2005generic}%
  \BibitemOpen
  \bibfield  {author} {\bibinfo {author} {\bibfnamefont {K.}~\bibnamefont
  {Kruse}}, \bibinfo {author} {\bibfnamefont {J.-F.}\ \bibnamefont {Joanny}},
  \bibinfo {author} {\bibfnamefont {F.}~\bibnamefont {J{\"u}licher}}, \bibinfo
  {author} {\bibfnamefont {J.}~\bibnamefont {Prost}},\ and\ \bibinfo {author}
  {\bibfnamefont {K.}~\bibnamefont {Sekimoto}},\ }\href@noop {} {\bibfield
  {journal} {\bibinfo  {journal} {The European Physical Journal E}\ }\textbf
  {\bibinfo {volume} {16}},\ \bibinfo {pages} {5} (\bibinfo {year}
  {2005})}\BibitemShut {NoStop}%
\bibitem [{\citenamefont {Giomi}\ \emph {et~al.}(2013)\citenamefont {Giomi},
  \citenamefont {Bowick}, \citenamefont {Ma},\ and\ \citenamefont
  {Marchetti}}]{giomi2013defect}%
  \BibitemOpen
  \bibfield  {author} {\bibinfo {author} {\bibfnamefont {L.}~\bibnamefont
  {Giomi}}, \bibinfo {author} {\bibfnamefont {M.~J.}\ \bibnamefont {Bowick}},
  \bibinfo {author} {\bibfnamefont {X.}~\bibnamefont {Ma}},\ and\ \bibinfo
  {author} {\bibfnamefont {M.~C.}\ \bibnamefont {Marchetti}},\ }\href@noop {}
  {\bibfield  {journal} {\bibinfo  {journal} {Physical review letters}\
  }\textbf {\bibinfo {volume} {110}},\ \bibinfo {pages} {228101} (\bibinfo
  {year} {2013})}\BibitemShut {NoStop}%
\bibitem [{\citenamefont {Saw}\ \emph {et~al.}(2017)\citenamefont {Saw},
  \citenamefont {Doostmohammadi}, \citenamefont {Nier}, \citenamefont
  {Kocgozlu}, \citenamefont {Thampi}, \citenamefont {Toyama}, \citenamefont
  {Marcq}, \citenamefont {Lim}, \citenamefont {Yeomans},\ and\ \citenamefont
  {Ladoux}}]{saw2017topological}%
  \BibitemOpen
  \bibfield  {author} {\bibinfo {author} {\bibfnamefont {T.~B.}\ \bibnamefont
  {Saw}}, \bibinfo {author} {\bibfnamefont {A.}~\bibnamefont {Doostmohammadi}},
  \bibinfo {author} {\bibfnamefont {V.}~\bibnamefont {Nier}}, \bibinfo {author}
  {\bibfnamefont {L.}~\bibnamefont {Kocgozlu}}, \bibinfo {author}
  {\bibfnamefont {S.}~\bibnamefont {Thampi}}, \bibinfo {author} {\bibfnamefont
  {Y.}~\bibnamefont {Toyama}}, \bibinfo {author} {\bibfnamefont
  {P.}~\bibnamefont {Marcq}}, \bibinfo {author} {\bibfnamefont {C.~T.}\
  \bibnamefont {Lim}}, \bibinfo {author} {\bibfnamefont {J.~M.}\ \bibnamefont
  {Yeomans}},\ and\ \bibinfo {author} {\bibfnamefont {B.}~\bibnamefont
  {Ladoux}},\ }\href@noop {} {\bibfield  {journal} {\bibinfo  {journal}
  {Nature}\ }\textbf {\bibinfo {volume} {544}},\ \bibinfo {pages} {212}
  (\bibinfo {year} {2017})}\BibitemShut {NoStop}%
\bibitem [{\citenamefont {Streichan}\ \emph {et~al.}(2018)\citenamefont
  {Streichan}, \citenamefont {Lefebvre}, \citenamefont {Noll}, \citenamefont
  {Wieschaus},\ and\ \citenamefont {Shraiman}}]{streichan2018global}%
  \BibitemOpen
  \bibfield  {author} {\bibinfo {author} {\bibfnamefont {S.~J.}\ \bibnamefont
  {Streichan}}, \bibinfo {author} {\bibfnamefont {M.~F.}\ \bibnamefont
  {Lefebvre}}, \bibinfo {author} {\bibfnamefont {N.}~\bibnamefont {Noll}},
  \bibinfo {author} {\bibfnamefont {E.~F.}\ \bibnamefont {Wieschaus}},\ and\
  \bibinfo {author} {\bibfnamefont {B.~I.}\ \bibnamefont {Shraiman}},\
  }\href@noop {} {\bibfield  {journal} {\bibinfo  {journal} {Elife}\ }\textbf
  {\bibinfo {volume} {7}},\ \bibinfo {pages} {e27454} (\bibinfo {year}
  {2018})}\BibitemShut {NoStop}%
\bibitem [{\citenamefont {Serra}\ \emph {et~al.}(2021)\citenamefont {Serra},
  \citenamefont {N{\'a}jera}, \citenamefont {Chuai}, \citenamefont {Spandan},
  \citenamefont {Weijer},\ and\ \citenamefont
  {Mahadevan}}]{serra2021mechanochemical}%
  \BibitemOpen
  \bibfield  {author} {\bibinfo {author} {\bibfnamefont {M.}~\bibnamefont
  {Serra}}, \bibinfo {author} {\bibfnamefont {G.~S.}\ \bibnamefont
  {N{\'a}jera}}, \bibinfo {author} {\bibfnamefont {M.}~\bibnamefont {Chuai}},
  \bibinfo {author} {\bibfnamefont {V.}~\bibnamefont {Spandan}}, \bibinfo
  {author} {\bibfnamefont {C.~J.}\ \bibnamefont {Weijer}},\ and\ \bibinfo
  {author} {\bibfnamefont {L.}~\bibnamefont {Mahadevan}},\ }\href@noop {}
  {\bibfield  {journal} {\bibinfo  {journal} {bioRxiv}\ ,\ \bibinfo {pages}
  {2021}} (\bibinfo {year} {2021})}\BibitemShut {NoStop}%
\bibitem [{\citenamefont {Chuai}\ \emph {et~al.}(2023)\citenamefont {Chuai},
  \citenamefont {Serrano~N{\'a}jera}, \citenamefont {Serra}, \citenamefont
  {Mahadevan},\ and\ \citenamefont {Weijer}}]{chuai2023reconstruction}%
  \BibitemOpen
  \bibfield  {author} {\bibinfo {author} {\bibfnamefont {M.}~\bibnamefont
  {Chuai}}, \bibinfo {author} {\bibfnamefont {G.}~\bibnamefont
  {Serrano~N{\'a}jera}}, \bibinfo {author} {\bibfnamefont {M.}~\bibnamefont
  {Serra}}, \bibinfo {author} {\bibfnamefont {L.}~\bibnamefont {Mahadevan}},\
  and\ \bibinfo {author} {\bibfnamefont {C.~J.}\ \bibnamefont {Weijer}},\
  }\href@noop {} {\bibfield  {journal} {\bibinfo  {journal} {Science Advances}\
  }\textbf {\bibinfo {volume} {9}},\ \bibinfo {pages} {eabn5429} (\bibinfo
  {year} {2023})}\BibitemShut {NoStop}%
\bibitem [{\citenamefont {Ibrahimi}\ and\ \citenamefont
  {Merkel}(2022)}]{ibrahimi2022deforming}%
  \BibitemOpen
  \bibfield  {author} {\bibinfo {author} {\bibfnamefont {M.}~\bibnamefont
  {Ibrahimi}}\ and\ \bibinfo {author} {\bibfnamefont {M.}~\bibnamefont
  {Merkel}},\ }\href@noop {} {\bibfield  {journal} {\bibinfo  {journal} {New
  Journal of Physics}\ } (\bibinfo {year} {2022})}\BibitemShut {NoStop}%
\bibitem [{\citenamefont {Sokurenko}\ \emph {et~al.}(2008)\citenamefont
  {Sokurenko}, \citenamefont {Vogel},\ and\ \citenamefont
  {Thomas}}]{sokurenko2008catch}%
  \BibitemOpen
  \bibfield  {author} {\bibinfo {author} {\bibfnamefont {E.~V.}\ \bibnamefont
  {Sokurenko}}, \bibinfo {author} {\bibfnamefont {V.}~\bibnamefont {Vogel}},\
  and\ \bibinfo {author} {\bibfnamefont {W.~E.}\ \bibnamefont {Thomas}},\
  }\href@noop {} {\bibfield  {journal} {\bibinfo  {journal} {Cell host \&
  microbe}\ }\textbf {\bibinfo {volume} {4}},\ \bibinfo {pages} {314} (\bibinfo
  {year} {2008})}\BibitemShut {NoStop}%
\bibitem [{\citenamefont {Veigel}\ \emph {et~al.}(2003)\citenamefont {Veigel},
  \citenamefont {Molloy}, \citenamefont {Schmitz},\ and\ \citenamefont
  {Kendrick-Jones}}]{veigel2003load}%
  \BibitemOpen
  \bibfield  {author} {\bibinfo {author} {\bibfnamefont {C.}~\bibnamefont
  {Veigel}}, \bibinfo {author} {\bibfnamefont {J.~E.}\ \bibnamefont {Molloy}},
  \bibinfo {author} {\bibfnamefont {S.}~\bibnamefont {Schmitz}},\ and\ \bibinfo
  {author} {\bibfnamefont {J.}~\bibnamefont {Kendrick-Jones}},\ }\href@noop {}
  {\bibfield  {journal} {\bibinfo  {journal} {Nature cell biology}\ }\textbf
  {\bibinfo {volume} {5}},\ \bibinfo {pages} {980} (\bibinfo {year}
  {2003})}\BibitemShut {NoStop}%
\bibitem [{\citenamefont {Sknepnek}\ \emph {et~al.}(2021)\citenamefont
  {Sknepnek}, \citenamefont {Henkes}, \citenamefont {Djafer-Cherif},\ and\
  \citenamefont {Weijer}}]{sknepnek2021generating}%
  \BibitemOpen
  \bibfield  {author} {\bibinfo {author} {\bibfnamefont {R.}~\bibnamefont
  {Sknepnek}}, \bibinfo {author} {\bibfnamefont {S.}~\bibnamefont {Henkes}},
  \bibinfo {author} {\bibfnamefont {I.}~\bibnamefont {Djafer-Cherif}},\ and\
  \bibinfo {author} {\bibfnamefont {C.~J.}\ \bibnamefont {Weijer}},\
  }\href@noop {} {\bibfield  {journal} {\bibinfo  {journal} {arXiv preprint
  arXiv:2106.12394}\ } (\bibinfo {year} {2021})}\BibitemShut {NoStop}%
\bibitem [{\citenamefont {Cl{\'e}ment}\ \emph {et~al.}(2017)\citenamefont
  {Cl{\'e}ment}, \citenamefont {Dehapiot}, \citenamefont {Collinet},
  \citenamefont {Lecuit},\ and\ \citenamefont
  {Lenne}}]{clement2017viscoelastic}%
  \BibitemOpen
  \bibfield  {author} {\bibinfo {author} {\bibfnamefont {R.}~\bibnamefont
  {Cl{\'e}ment}}, \bibinfo {author} {\bibfnamefont {B.}~\bibnamefont
  {Dehapiot}}, \bibinfo {author} {\bibfnamefont {C.}~\bibnamefont {Collinet}},
  \bibinfo {author} {\bibfnamefont {T.}~\bibnamefont {Lecuit}},\ and\ \bibinfo
  {author} {\bibfnamefont {P.-F.}\ \bibnamefont {Lenne}},\ }\href@noop {}
  {\bibfield  {journal} {\bibinfo  {journal} {Current biology}\ }\textbf
  {\bibinfo {volume} {27}},\ \bibinfo {pages} {3132} (\bibinfo {year}
  {2017})}\BibitemShut {NoStop}%
\bibitem [{\citenamefont {Khalilgharibi}\ \emph {et~al.}(2019)\citenamefont
  {Khalilgharibi}, \citenamefont {Fouchard}, \citenamefont {Asadipour},
  \citenamefont {Barrientos}, \citenamefont {Duda}, \citenamefont {Bonfanti},
  \citenamefont {Yonis}, \citenamefont {Harris}, \citenamefont {Mosaffa},
  \citenamefont {Fujita} \emph {et~al.}}]{khalilgharibi2019stress}%
  \BibitemOpen
  \bibfield  {author} {\bibinfo {author} {\bibfnamefont {N.}~\bibnamefont
  {Khalilgharibi}}, \bibinfo {author} {\bibfnamefont {J.}~\bibnamefont
  {Fouchard}}, \bibinfo {author} {\bibfnamefont {N.}~\bibnamefont {Asadipour}},
  \bibinfo {author} {\bibfnamefont {R.}~\bibnamefont {Barrientos}}, \bibinfo
  {author} {\bibfnamefont {M.}~\bibnamefont {Duda}}, \bibinfo {author}
  {\bibfnamefont {A.}~\bibnamefont {Bonfanti}}, \bibinfo {author}
  {\bibfnamefont {A.}~\bibnamefont {Yonis}}, \bibinfo {author} {\bibfnamefont
  {A.}~\bibnamefont {Harris}}, \bibinfo {author} {\bibfnamefont
  {P.}~\bibnamefont {Mosaffa}}, \bibinfo {author} {\bibfnamefont
  {Y.}~\bibnamefont {Fujita}}, \emph {et~al.},\ }\href@noop {} {\bibfield
  {journal} {\bibinfo  {journal} {Nature physics}\ }\textbf {\bibinfo {volume}
  {15}},\ \bibinfo {pages} {839} (\bibinfo {year} {2019})}\BibitemShut
  {NoStop}%
\bibitem [{\citenamefont {Scheibner}\ \emph {et~al.}(2020)\citenamefont
  {Scheibner}, \citenamefont {Souslov}, \citenamefont {Banerjee}, \citenamefont
  {Sur{\'o}wka}, \citenamefont {Irvine},\ and\ \citenamefont
  {Vitelli}}]{scheibner2020odd}%
  \BibitemOpen
  \bibfield  {author} {\bibinfo {author} {\bibfnamefont {C.}~\bibnamefont
  {Scheibner}}, \bibinfo {author} {\bibfnamefont {A.}~\bibnamefont {Souslov}},
  \bibinfo {author} {\bibfnamefont {D.}~\bibnamefont {Banerjee}}, \bibinfo
  {author} {\bibfnamefont {P.}~\bibnamefont {Sur{\'o}wka}}, \bibinfo {author}
  {\bibfnamefont {W.~T.}\ \bibnamefont {Irvine}},\ and\ \bibinfo {author}
  {\bibfnamefont {V.}~\bibnamefont {Vitelli}},\ }\href@noop {} {\bibfield
  {journal} {\bibinfo  {journal} {Nature Physics}\ }\textbf {\bibinfo {volume}
  {16}},\ \bibinfo {pages} {475} (\bibinfo {year} {2020})}\BibitemShut
  {NoStop}%
\bibitem [{\citenamefont {Brandenbourger}\ \emph {et~al.}(2021)\citenamefont
  {Brandenbourger}, \citenamefont {Scheibner}, \citenamefont {Veenstra},
  \citenamefont {Vitelli},\ and\ \citenamefont
  {Coulais}}]{brandenbourger2021limit}%
  \BibitemOpen
  \bibfield  {author} {\bibinfo {author} {\bibfnamefont {M.}~\bibnamefont
  {Brandenbourger}}, \bibinfo {author} {\bibfnamefont {C.}~\bibnamefont
  {Scheibner}}, \bibinfo {author} {\bibfnamefont {J.}~\bibnamefont {Veenstra}},
  \bibinfo {author} {\bibfnamefont {V.}~\bibnamefont {Vitelli}},\ and\ \bibinfo
  {author} {\bibfnamefont {C.}~\bibnamefont {Coulais}},\ }\href@noop {}
  {\bibfield  {journal} {\bibinfo  {journal} {arXiv preprint arXiv:2108.08837}\
  } (\bibinfo {year} {2021})}\BibitemShut {NoStop}%
\bibitem [{\citenamefont {Banerjee}\ \emph {et~al.}(2021)\citenamefont
  {Banerjee}, \citenamefont {Vitelli}, \citenamefont {J{\"u}licher},\ and\
  \citenamefont {Sur{\'o}wka}}]{banerjee2021active}%
  \BibitemOpen
  \bibfield  {author} {\bibinfo {author} {\bibfnamefont {D.}~\bibnamefont
  {Banerjee}}, \bibinfo {author} {\bibfnamefont {V.}~\bibnamefont {Vitelli}},
  \bibinfo {author} {\bibfnamefont {F.}~\bibnamefont {J{\"u}licher}},\ and\
  \bibinfo {author} {\bibfnamefont {P.}~\bibnamefont {Sur{\'o}wka}},\
  }\href@noop {} {\bibfield  {journal} {\bibinfo  {journal} {Physical Review
  Letters}\ }\textbf {\bibinfo {volume} {126}},\ \bibinfo {pages} {138001}
  (\bibinfo {year} {2021})}\BibitemShut {NoStop}%
\end{thebibliography}%

\clearpage
\newpage

\begin{widetext}
\begin{center}
\textbf{\large Mechanochemical active feedback generates convergent-extension in a continuum model for epithelial tissue: Supplemental Material}
\end{center}
\end{widetext}

\setcounter{equation}{0}
\setcounter{figure}{0}
\setcounter{table}{0}
\setcounter{section}{0}
\setcounter{page}{1}
\makeatletter
\renewcommand{\theequation}{S\arabic{equation}}
\renewcommand{\thefigure}{S\arabic{figure}}
\renewcommand{\bibnumfmt}[1]{[S#1]}
\renewcommand{\citenumfont}[1]{S#1}

\onecolumngrid

\section{Additional data} \label{sec_additional_data}

\begin{figure*}[h]
	\centering
	\includegraphics[width = 1.0\textwidth]{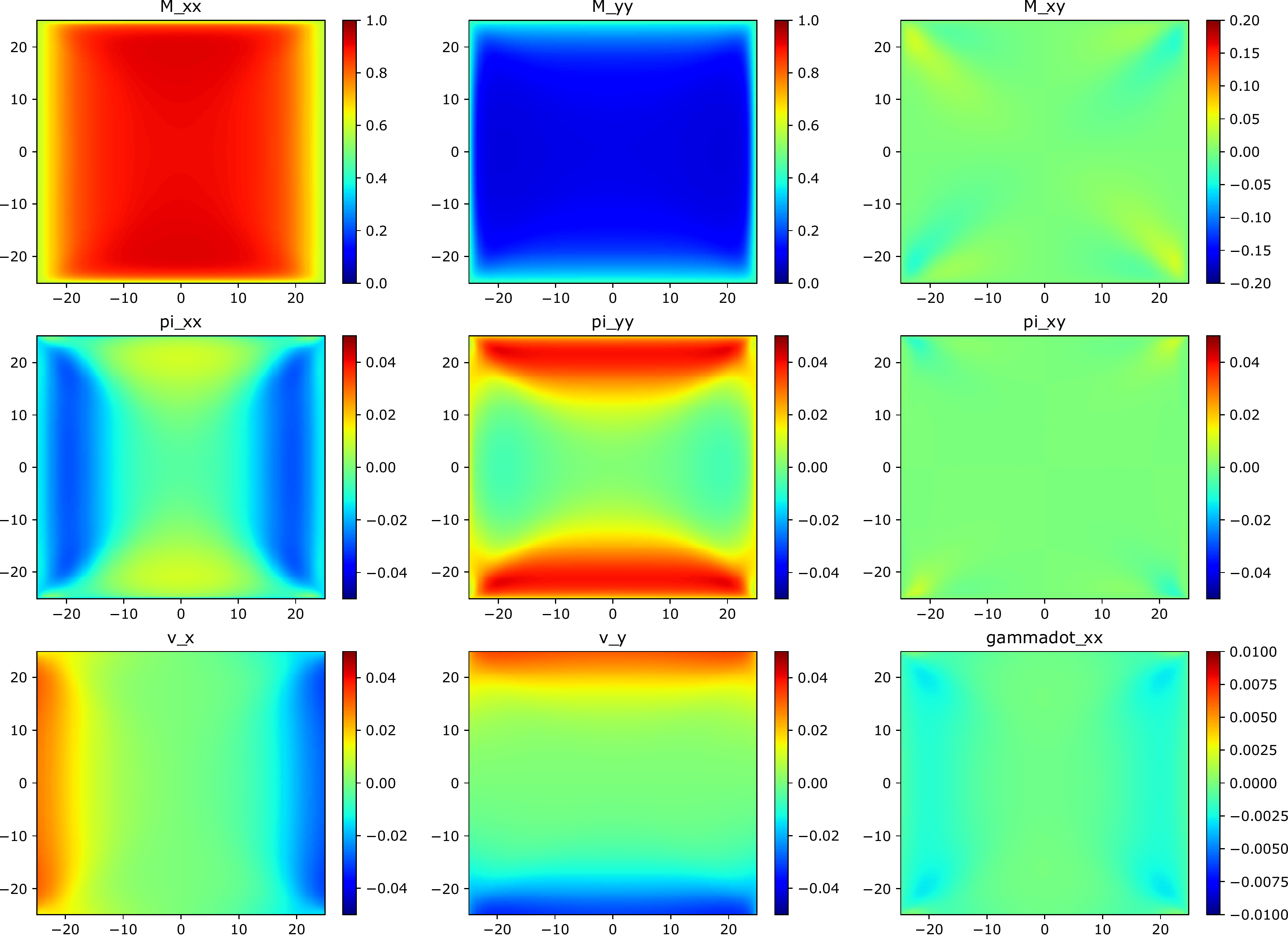}
	\caption{Convergence extension phase with parameters $\beta = 0.7$, $\sigma^{\text{ext}}_{xx} = 0.04$, $\sigma^{\text{ext}}_{yy} = -0.04$, $\tau_m = 20.0$, $\tau_v = 20.0$. Tension is applied to the patch of the tissue parallel to the $x$ axis, and compression parallel to the $y$ axis. The top row has the components of the ActoMyosin tensor: $M_{xx}, M_{yy}, M_{xy}$, the middle row has the components of the passive stress tensor: $\pi_{xx}, \pi_{yy}, \pi_{xy}$, and the bottom row has components of velocity and the $xx$ component of the strain rate tensor: $v_x, v_y, \dot{\gamma}_{xx}$.}  
\label{sfig:figure_1}
\end{figure*}

\pagebreak

\begin{figure*}[h]
	\centering
	\includegraphics[width = 1.0\textwidth]{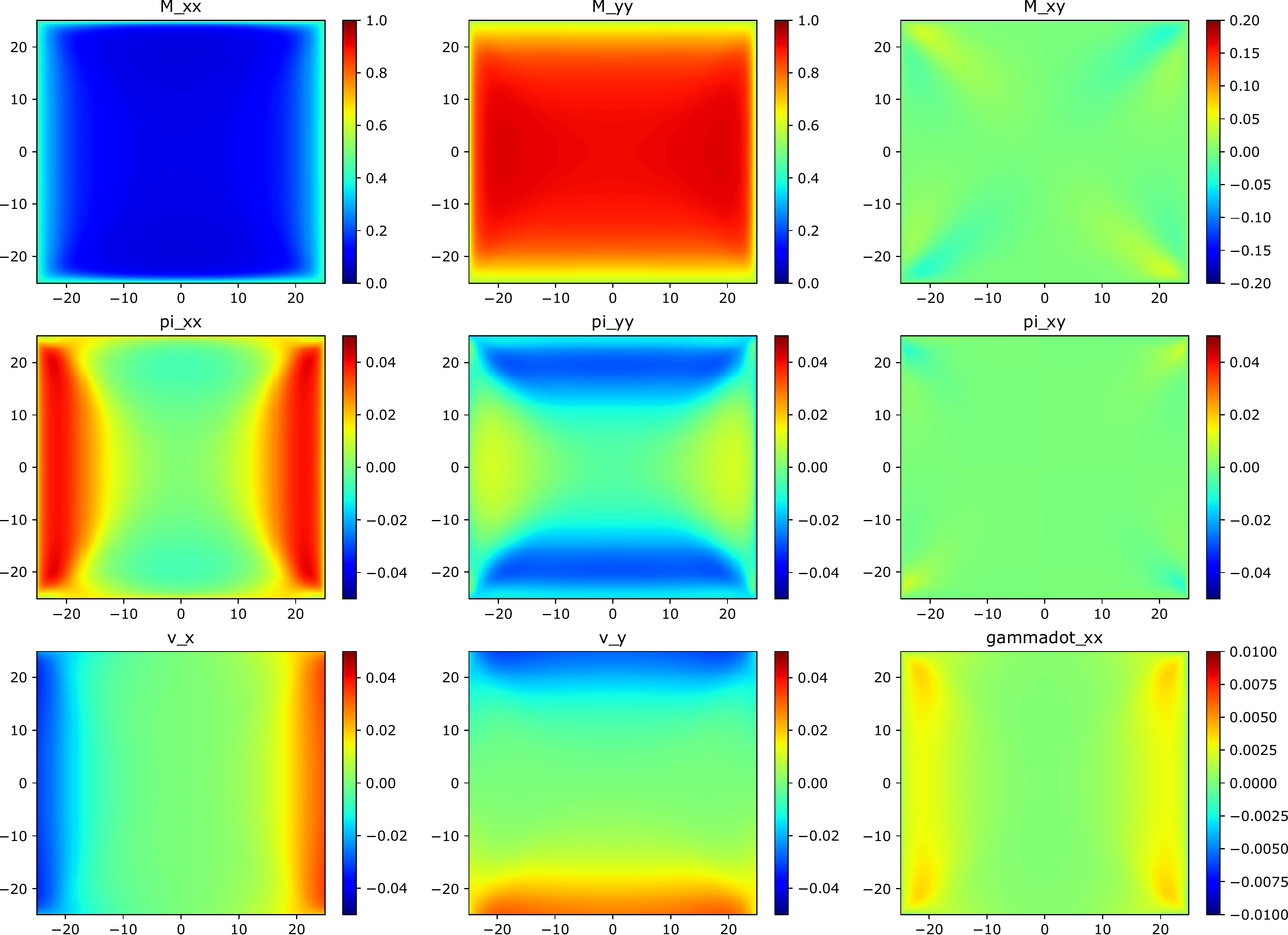}
	\caption{Convergence extension phase with parameters $\beta = 0.7$, $\sigma^{\text{ext}}_{xx} = -0.04$, $\sigma^{\text{ext}}_{yy} = 0.04$, $\tau_m = 20.0$, $\tau_v = 20.0$. Compression is applied to the patch of the tissue parallel to the $x$ axis, and tension parallel to the $y$ axis. The top row has the components of the ActoMyosin tensor: $M_{xx}, M_{yy}, M_{xy}$, the middle row has the components of the passive stress tensor: $\pi_{xx}, \pi_{yy}, \pi_{xy}$, and the bottom row has components of velocity and the $xx$ component of the strain rate tensor: $v_x, v_y, \dot{\gamma}_{xx}$.}  
\label{sfig:figure_2}
\end{figure*}

\pagebreak

\begin{figure*}[h]
	\centering
	\includegraphics[width = 1.0\textwidth]{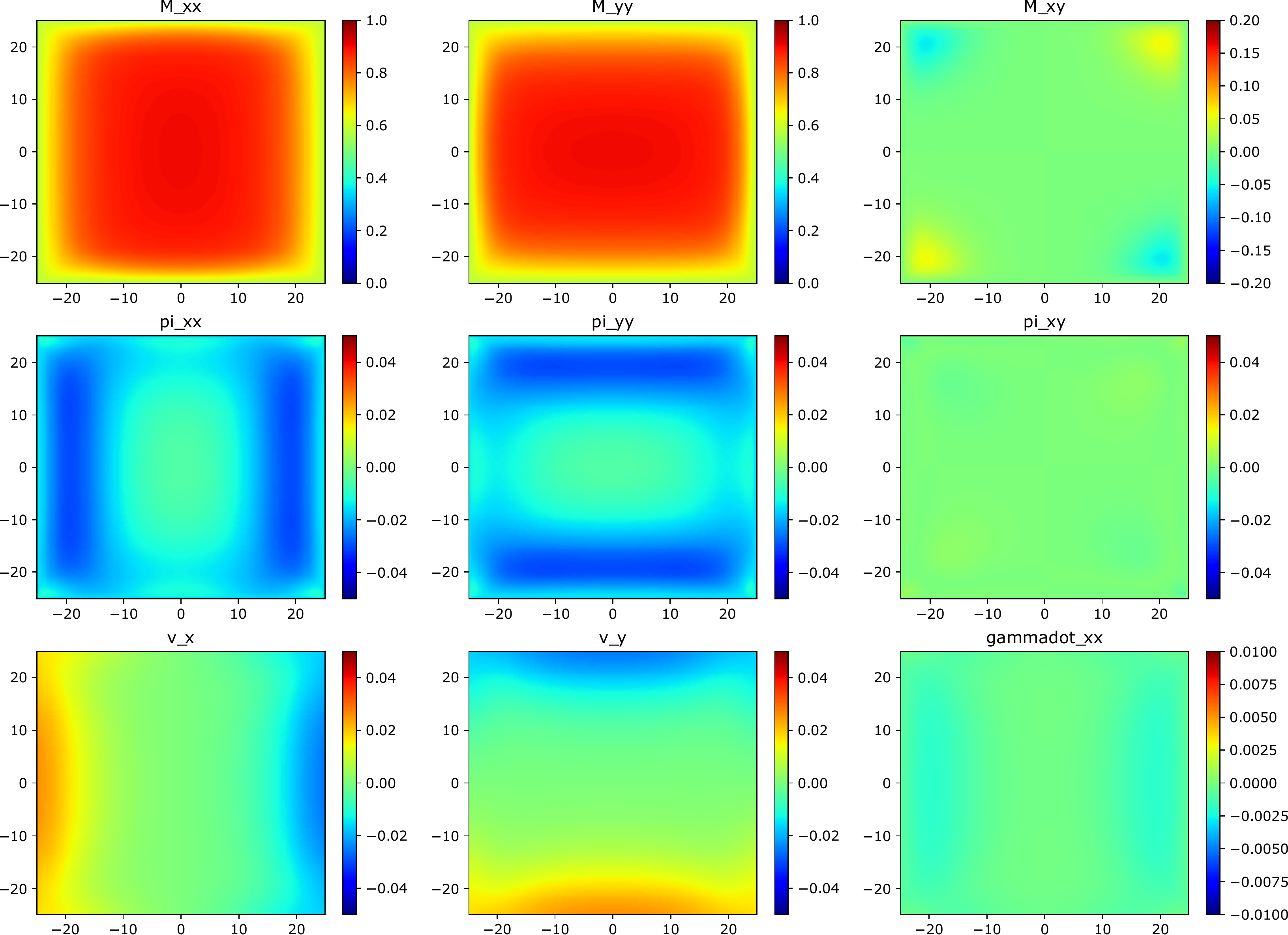}
	\caption{Convergence extension phase with parameters $\beta = 0.7$, $\sigma^{\text{ext}}_{xx} = 0.04$, $\sigma^{\text{ext}}_{yy} = 0.04$, $\tau_m = 20.0$, $\tau_v = 20.0$. The patch of tissue is being pulled uniformly. The top row has the components of the ActoMyosin tensor: $M_{xx}, M_{yy}, M_{xy}$, the middle row has the components of the passive stress tensor: $\pi_{xx}, \pi_{yy}, \pi_{xy}$, and the bottom row has components of velocity and the $xx$ component of the strain rate tensor: $v_x, v_y, \dot{\gamma}_{xx}$.}  
\label{sfig:figure_3}
\end{figure*}

\pagebreak

\begin{figure*}[h]
	\centering
	\includegraphics[width = 1.0\textwidth]{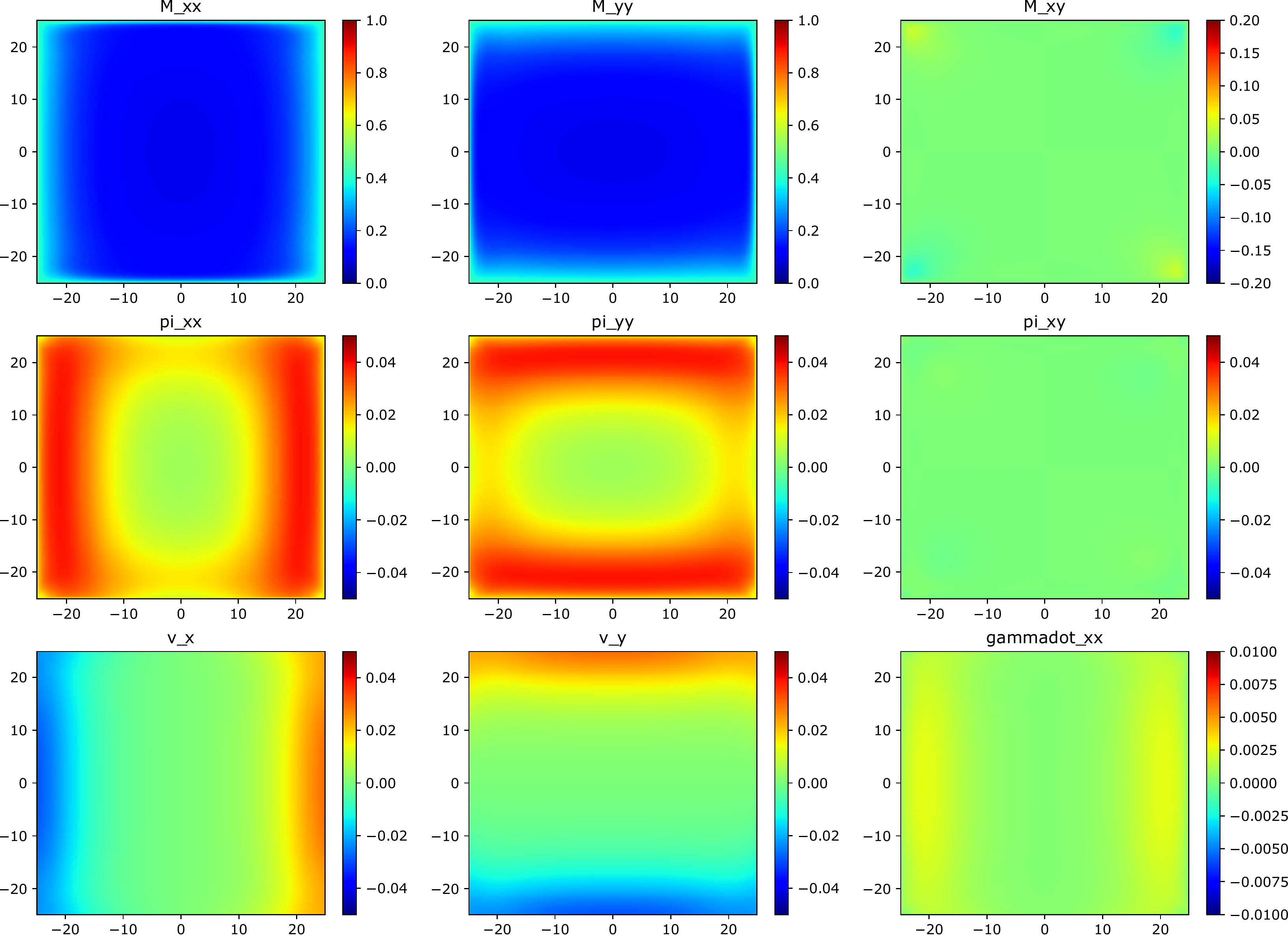}
	\caption{Convergence extension phase with parameters $\beta = 0.7$, $\sigma^{\text{ext}}_{xx} = -0.04$, $\sigma^{\text{ext}}_{yy} = -0.04$, $\tau_m = 20.0$, $\tau_v = 20.0$. The patch of tissue is being compressed uniformly. The top row has the components of the ActoMyosin tensor: $M_{xx}, M_{yy}, M_{xy}$, the middle row has the components of the passive stress tensor: $\pi_{xx}, \pi_{yy}, \pi_{xy}$, and the bottom row has components of velocity and the $xx$ component of the strain rate tensor: $v_x, v_y, \dot{\gamma}_{xx}$.}  
\label{sfig:figure_4}
\end{figure*}

\pagebreak

\begin{figure*}[h]
	\centering
	\includegraphics[width = 1.0\textwidth]{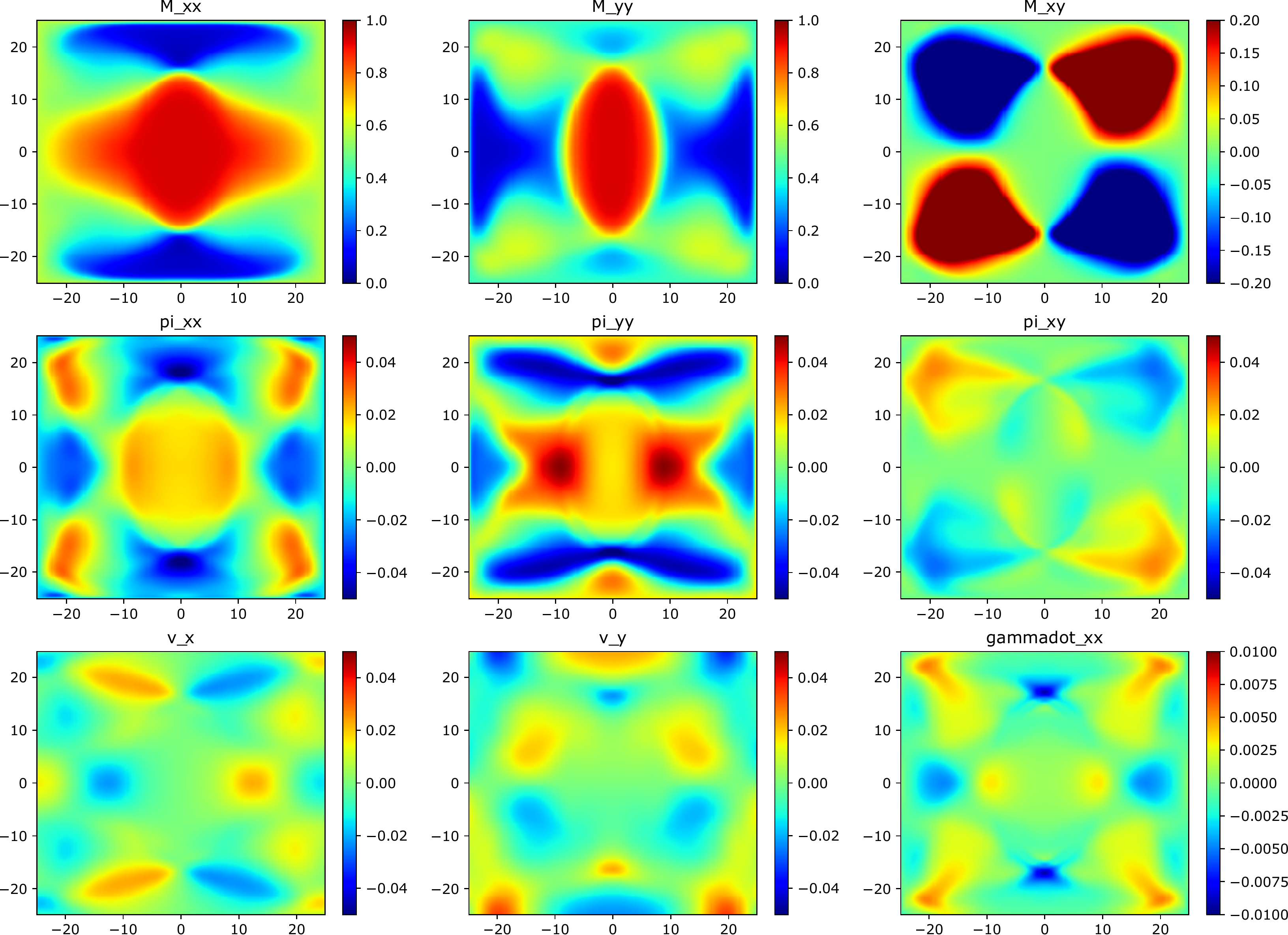}
	\caption{Oscillating phase with parameters $\beta = 0.7$, $\sigma^{\text{ext}}_{xx} = 0.04$, $\sigma^{\text{ext}}_{yy} = -0.04$, $\tau_m = 5.0$, $\tau_v = 40.0$. Tension is applied to the patch of the tissue parallel to the $x$ axis, and compression parallel to the $y$ axis. The top row has the components of the ActoMyosin tensor: $M_{xx}, M_{yy}, M_{xy}$, the middle row has the components of the passive stress tensor: $\pi_{xx}, \pi_{yy}, \pi_{xy}$, and the bottom row has components of velocity and the $xx$ component of the strain rate tensor: $v_x, v_y, \dot{\gamma}_{xx}$.}  
\label{sfig:figure_5}
\end{figure*}

\pagebreak

\begin{figure*}[h]
	\centering
	\includegraphics[width = 1.0\textwidth]{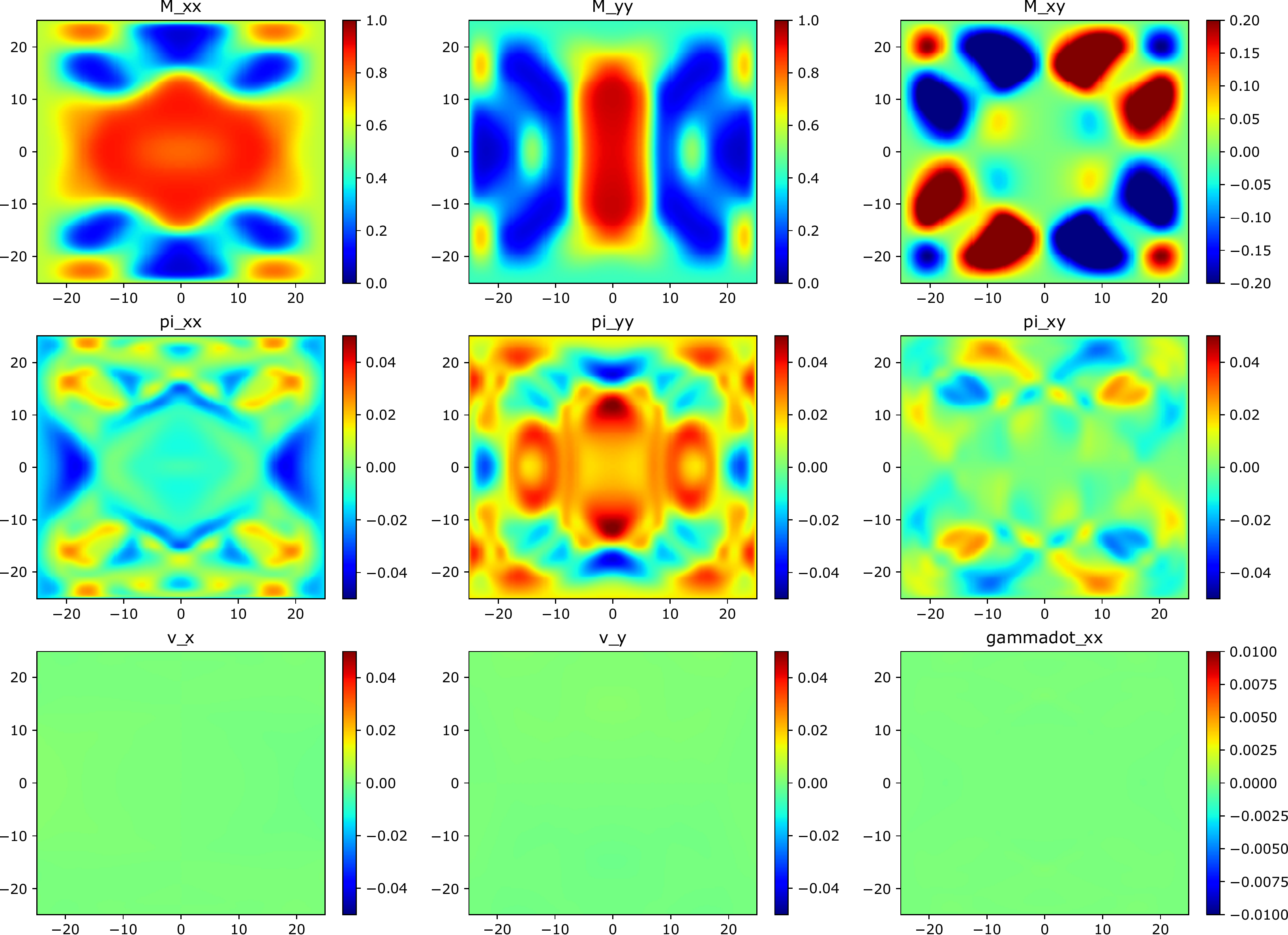}
	\caption{Complex spatial pattern phase with parameters $\beta = 0.7$, $\sigma^{\text{ext}}_{xx} = 0.04$, $\sigma^{\text{ext}}_{yy} = -0.04$, $\tau_m = 200.0$, $\tau_v = 500.0$. Tension is applied to the patch of the tissue parallel to the $x$ axis, and compression parallel to the $y$ axis. The top row has the components of the ActoMyosin tensor: $M_{xx}, M_{yy}, M_{xy}$, the middle row has the components of the passive stress tensor: $\pi_{xx}, \pi_{yy}, \pi_{xy}$, and the bottom row has components of velocity and the $xx$ component of the strain rate tensor: $v_x, v_y, \dot{\gamma}_{xx}$.}  
\label{sfig:figure_6}
\end{figure*}

\pagebreak

\begin{figure*}[h]
	\centering
	\includegraphics[width = 1.0\textwidth]{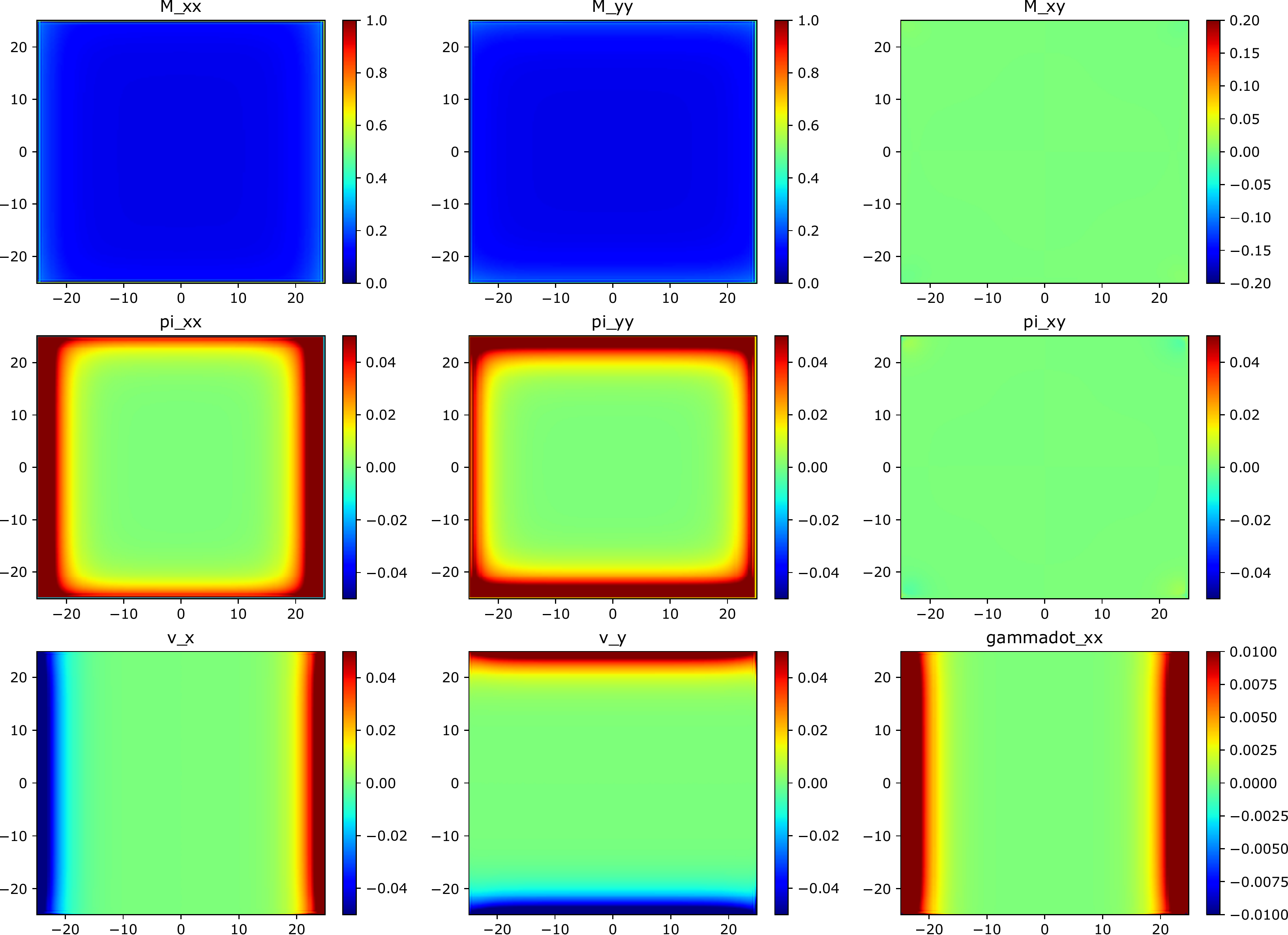}
	\caption{Local expansion phase with parameters $\beta = 0.7$, $\sigma^{\text{ext}}_{xx} = 0.04$, $\sigma^{\text{ext}}_{yy} = -0.04$, $\tau_m = 20.0$, $\tau_v = 20.0$. Tension is applied to the patch of the tissue parallel to the $x$ axis, and compression parallel to the $y$ axis. The top row has the components of the ActoMyosin tensor: $M_{xx}, M_{yy}, M_{xy}$, the middle row has the components of the passive stress tensor: $\pi_{xx}, \pi_{yy}, \pi_{xy}$, and the bottom row has components of velocity and the $xx$ component of the strain rate tensor: $v_x, v_y, \dot{\gamma}_{xx}$.}  
\label{sfig:figure_7}
\end{figure*}

\pagebreak

\begin{figure*}[h]
	\centering
	\includegraphics[width = 0.35\textwidth]{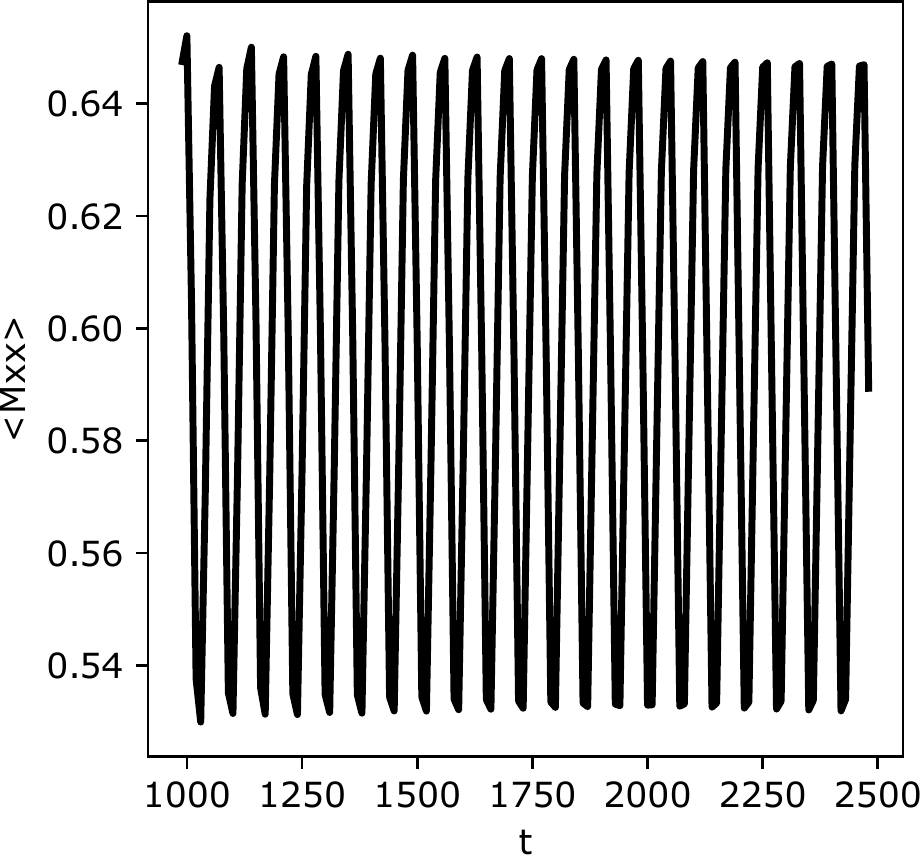}
	\caption{Spatial average of $M_{xx}$ as a function of simulation time for an oscillating phase with parameters $\beta = 0.7$, $\sigma_{\text{ext}} = 0.04$, $\tau_m = 5.0$, $\tau_v = 40.0$.}  
\label{sfig:figure_8}
\end{figure*}

\section{Single junction model} \label{sec_myo_eqn}

The mechano-chemical feedback in the ActoMyosin equation is inspired by a model for a contractile active junction wirtten down by SH and colleagues. Let us consider a single junction of length $l$ and rest length $l_0$. The junction is contractile and has myosin concentration $m$, and is subject to an external pulling force $\sigma_{ext}$. Assuming the passive mechanical and active component of the junction are acting in parallel, the equation of motion is
\begin{equation}\label{eq:single_junc_eom}
\zeta\dot{l} = -k(l - l_0) - \beta(m - m_0) + \sigma_{ext},
\end{equation}
where $\beta$ is the activity, $\zeta$ is the friction between the junction and the surrounding medium, $k$ is the stiffness of the junction, and $m_0$ is the reference value for $m$. The rest length $l_0$ of the junction undergoes viscous relaxation
\begin{equation} \label{eq:single_junc_relaxation}
\lambda\dot{l}_0 = -(l_0 - l),
\end{equation}
where $\lambda$ is the viscous relaxation time of the junction. Finally, we include active feedback by making the unbinding rate of myosin decrease with tension
\begin{equation} \label{eq:single_junc_myosin}
\tau_m \dot{m} = 1 - m f(\sigma),
\end{equation}
where $\tau_m^{-1}$ is the myosin binding rate, $f(\sigma)/\tau_m$ is the myosin unbinding rate, and one-dimensional stress $\sigma = k(l - l_0) + \beta(m - m_0)$. The myosin unbinding rate decreases exponentially with tension as
\begin{equation} \label{eq:single_junc_feedback_func}
f(\sigma) = 1 + e^{-k_0\sigma}.
\end{equation}

\section{Approximate solution for convergence extension} \label{sec_aprrox_soln}

We have derived an approximate solution for the convergence extension steady state by expanding around the fixed points $\pi_{\alpha \alpha}(M_{\alpha \alpha}) = 0$ of the mean field to linear order. We write
\begin{subequations}
\begin{align}
M_{xx}(\bm{r},t) & = m^{+} + m_x(\bm{r}, t), \\ M_{yy}(\bm{r},t) & = m^{-} + m_y(\bm{r}, t),
\end{align}
\end{subequations}
and expand
\begin{subequations}
\begin{align}
\pi_{xx}(M_{xx}(\bm{r},t)) & = \pi_{xx}^{\prime}(m^{+})m_x(\bm{r},t), \\ 
\pi_{yy}(M_{yy}(\bm{r},t)) & = \pi_{yy}^{\prime}(m^{-})m_y(\bm{r},t),
\end{align}
\end{subequations}
where we have used the fact that $\pi_{\alpha \alpha}(m^{\pm}) = 0$. From equation \eqref{eq:momentum_balance}, the velocity is given by
\begin{equation}
\begin{aligned}
v_x & = \bigg(\frac{\pi_{xx}^{\prime}(m^+) + \beta}{\zeta}\bigg)\partial_x m_x, \\
v_y & = \bigg(\frac{\pi_{yy}^{\prime}(m^-) + \beta}{\zeta}\bigg)\partial_y m_y.
\end{aligned}
\end{equation}
 We set the off diagonal components of active and passive stress equal to zero. We look at times long compared to the myosin unbinding time $\tau_m$. We write equation \eqref{eq:compressible_maxwell} to linear order in $m_x$ and $m_y$, which means that we can ignore the term ${\bf{v}}\cdot\nabla  \pi_{\alpha\alpha}$ because its lowest order terms are proportional to $m_x^2, m_y^2, m_xm_y$. The other terms in the corotational derivative vanish because the off diagonal components of the passive stress are zero. This leaves us with 
\begin{subequations}
\begin{align}
\pi_{xx}^{\prime}(m^{+})(m_x + \lambda\partial_t m_x) & = A_{+}\partial^2_xm_x + A_{-}\partial^2_ym_y, \label{eq:si_linear_passive_xx}\\
\pi_{yy}^{\prime}(m^{-})(m_y + \lambda\partial_t m_y) & = B_{+}\partial^2_ym_y + B_{-}\partial^2_xm_x, \label{eq:si_linear_passive_yy}\\
0 & = \partial_x\partial_y(C_+m_x + C_-m_y), \label{eq:si_linear_passive_xy}
\end{align}
\end{subequations}
where we have used equation \eqref{eq:momentum_balance} to write the strain rate, and where 
\begin{align}
& A_+ = \frac{\eta_p + \eta_s}{2\zeta}(\pi_{xx}^{\prime}(m^+) + \beta), && A_- = \frac{\eta_p - \eta_s}{2\zeta}(\pi_{yy}^{\prime}(m^-) + \beta), \nonumber \\ 
& B_+ = \frac{\eta_p + \eta_s}{2\zeta}(\pi_{yy}^{\prime}(m^-) + \beta), && B_- = \frac{\eta_p - \eta_s}{2\zeta}(\pi_{xx}^{\prime}(m^+) + \beta), \\ 
& C_+ = \frac{\eta_s}{2\zeta}(\pi_{xx}^{\prime}(m^+) + \beta), && C_- = \frac{\eta_s}{2\zeta}(\pi_{yy}^{\prime}(m^-) + \beta). \nonumber
\end{align}
If we work in the limit $t \gg \lambda$, then we can neglect the time derivative since $\lambda\partial_t m_x, \lambda\partial_t m_y \to 0$. This gives coupled partial differential equations in $x$ and $y$ for $m_x(x,y)$ and $m_y(x,y)$. Equation \eqref{eq:si_linear_passive_xy} suggests that we looks for solutions of the form $m_x(x,y) = X_{1}(x) + Y_{1}(y)$, $m_x(x,y) = X_{2}(x) + Y_{2}(y)$. Plugging this into \eqref{eq:si_linear_passive_xx} and \eqref{eq:si_linear_passive_yy}, it is straightforward to show that
\begin{align*}
& X_{1}(x) \sim e^{\pm x/\Lambda_{+}}, && Y_{1}(y) \sim e^{\pm y/\Lambda_{\pm}}, \\ & X_2(x) \sim e^{\pm x/\Lambda_{\mp}}, && Y_2(y) \sim e^{\pm y/\Lambda_{-}},
\end{align*}
where 
\begin{subequations}
\begin{align}
\Lambda_+ & = \sqrt{\bigg(\frac{\eta_p + \eta_s}{2\zeta}\bigg)\bigg(\frac{\pi_{xx}^{\prime}(m^+) + \beta}{\pi_{xx}^{\prime}(m^+)}\bigg)}, \\ 
\Lambda_{\pm} & = \sqrt{\bigg(\frac{\eta_p - \eta_s}{2\zeta}\bigg)\bigg(\frac{\pi_{yy}^{\prime}(m^-) + \beta}{\pi_{xx}^{\prime}(m^+)}\bigg)}, \\
\Lambda_{\mp} & = \sqrt{\bigg(\frac{\eta_p - \eta_s}{2\zeta}\bigg)\bigg(\frac{\pi_{xx}^{\prime}(m^+) + \beta}{\pi_{yy}^{\prime}(m^-)}\bigg)}, \\
\Lambda_- & = \sqrt{\bigg(\frac{\eta_p + \eta_s}{2\zeta}\bigg)\bigg(\frac{\pi_{yy}^{\prime}(m^-) + \beta}{\pi_{yy}^{\prime}(m^-)}\bigg)}.
\end{align}
\end{subequations}
We approximate $\pi_{xx}^{\prime}(m^+)$ and $\pi_{yy}^{\prime}(m^-)$ by differentiating equation \eqref{eq:nullclines} with respect to $M_{\alpha \alpha}$ and evaluating at $m^+$ and $m^-$. This solution is not compatible with constant ActoMyosin on the boundary, however there is good agreement with simulations (figure \ref{fig:figure_3}C and \ref{fig:figure_3}D compared to \ref{fig:figure_2}C and \ref{fig:figure_2}D) if we apply the boundary conditions like so: $M_{xx}(\pm L/2, 0) = M_{xx}^B$, $M_{xx}(0, \pm L/2) = M_{xx}^B$, $M_{yy}(\pm L/2, 0) = M_{yy}^B$, and $M_{yy}(0, \pm L/2) = M_{yy}^B$, where $M_{xx}^B$ and $M_{yy}^B$ are the isotropic equilibrium components of of the ActoMyosin tensor consistent with our chosen value $\bm{\sigma}^{\text{ext}}$ at the boundary:
\begin{align*}
M_{xx}^B = \frac{1}{1 + e^{-k_0\sigma_{\text{ext}}}}, && M_{yy}^B = \frac{1}{1 + e^{+k_0\sigma_{\text{ext}}}}.
\end{align*}
We find that
\begin{subequations} \label{si_full_approx_soln}
\begin{align}
m_x(x,y) & = \alpha_{xx}\bigg[\cosh\bigg(\frac{x}{\Lambda_+}\bigg)\bigg[1 - \cosh\bigg(\frac{L}{2\Lambda_{\pm}}\bigg)\bigg] + \cosh\bigg(\frac{y}{\Lambda_{\pm}}\bigg)\bigg[1 - \cosh\bigg(\frac{L}{2\Lambda_+}\bigg)\bigg]\bigg], \\ 
m_y(x,y) & = \alpha_{yy}\bigg[\cosh\bigg(\frac{x}{\Lambda_{\mp}}\bigg)\bigg[1 - \cosh\bigg(\frac{L}{2\Lambda_{-}}\bigg)\bigg] + \cosh\bigg(\frac{y}{\Lambda_{-}}\bigg)\bigg[1 - \cosh\bigg(\frac{L}{2\Lambda_{\mp}}\bigg)\bigg]\bigg],
\end{align}
\end{subequations}
where 
\begin{align*}
\alpha_{xx} = \frac{M_{xx}^B - m^+}{1 - \cosh\bigg(\frac{L}{2\Lambda_{+}}\bigg)\cosh\bigg(\frac{L}{2\Lambda_{\pm}}\bigg)}, && \alpha_{yy} = \frac{M_{yy}^B - m^-}{1 - \cosh\bigg(\frac{L}{2\Lambda_{-}}\bigg)\cosh\bigg(\frac{L}{2\Lambda_{\mp}}\bigg)}.
\end{align*}
The gradients of the passive stress are obtained by differentiating \eqref{eq:nullclines}
\begin{subequations}
\begin{align}
\pi_{xx}^{\prime}(m^+) & = \frac{1}{k_0m^+(1 - m^+)} - \beta, \\
\pi_{yy}^{\prime}(m^-) & = \frac{1}{k_0m^-(1 - m^-)} - \beta.
\end{align}
\end{subequations}
This solution depends on the viscous time scale via $\eta_p = \lambda B$ and $\eta_s = \lambda \mu$, where $B$ and $\mu$ are the bulk and shear moduli of the material. The velocities and strain rates are given by
\begin{subequations}
\begin{align}
v_x & = \frac{\alpha_{xx}}{\Lambda_{+}}\bigg(\frac{\pi_{xx}^{\prime}(m^+) + \beta}{\zeta}\bigg)\bigg[1 - \cosh\bigg(\frac{L}{2\Lambda_{\pm}}\bigg)\bigg]\sinh\bigg(\frac{x}{\Lambda_+}\bigg), \\
v_y & = \frac{\alpha_{yy}}{\Lambda_-}\bigg(\frac{\pi_{yy}^{\prime}(m^-) + \beta}{\zeta}\bigg)\bigg[1 - \cosh\bigg(\frac{L}{2\Lambda_{\mp}}\bigg)\bigg]\sinh\bigg(\frac{y}{\Lambda_{-}}\bigg), \\
\dot{\gamma}_{xx} & = \frac{\alpha_{xx}}{\Lambda_{+}^2}\bigg(\frac{\pi_{xx}^{\prime}(m^+) + \beta}{\zeta}\bigg)\bigg[1 - \cosh\bigg(\frac{L}{2\Lambda_{\pm}}\bigg)\bigg]\cosh\bigg(\frac{x}{\Lambda_+}\bigg), \\
\dot{\gamma}_{yy} & = \frac{\alpha_{yy}}{\Lambda_{-}^2}\bigg(\frac{\pi_{yy}^{\prime}(m^-) + \beta}{\zeta}\bigg)\bigg[1 - \cosh\bigg(\frac{L}{2\Lambda_{\mp}}\bigg)\bigg]\cosh\bigg(\frac{y}{\Lambda_{-}}\bigg).
\end{align}
\end{subequations}
We can calculate the length $\Lambda_d$ over which, say, $M_{yy}$ drops to $1/e$ of its boundary value via $M_{yy}(0,x_d) = e^{-1}M_{yy}(0,L/2)$, where $\Lambda_d = L/2 - x_d$. We find
\begin{equation}
x_d = \Lambda_-\cosh^{-1}\left[\frac{\cosh\bigg(\frac{L}{2\Lambda_-}\bigg) - 1 + \frac{1}{\alpha_{yy}}\bigg(e^{-1}\bigg[m^{-} + \alpha_{yy}\bigg(1 - \cosh\bigg(\frac{L}{2\Lambda_-}\bigg)\cosh\bigg(\frac{L}{2\lambda_{\pm}}\bigg)\bigg)\bigg] - m^{-}\bigg)}{1 - \cosh\bigg(\frac{L}{2\Lambda_{\mp}}\bigg)}\right].
\end{equation}
The spatially averaged pure shear strain is 
\begin{equation}
\begin{aligned}
\langle\dot{\gamma}_{xx} - \dot{\gamma}_{yy}\rangle & = \frac{2\alpha_{xx}}{L\Lambda_+}\bigg(\frac{\pi_{xx}^{\prime}(m^+) + \beta}{\zeta}\bigg)\bigg[1 - \cosh\bigg(\frac{L}{2\Lambda_{\pm}}\bigg)\bigg]\sinh\bigg(\frac{L}{2\Lambda_+}\bigg) \\ & - \frac{2\alpha_{yy}}{L\Lambda_-}\bigg(\frac{\pi_{yy}^{\prime}(m^-) + \beta}{\zeta}\bigg)\bigg[1 - \cosh\bigg(\frac{L}{2\Lambda_{\mp}}\bigg)\bigg]\sinh\bigg(\frac{L}{2\Lambda_{-}}\bigg).
\end{aligned}
\end{equation}

\section{Dynamics of the traceless part of the ActoMyosin tensor} \label{sec_active_nematics}
The ActoMyosin tensor can be written as the sum of the traceless part $\bm{Q}$ and a part proportional to the identity matrix
\begin{equation}
\bm{M} = 
\begin{pmatrix}
M_1 & M_2 \\ M_2 & -M_1
\end{pmatrix} + \frac{1}{2}\begin{pmatrix}
\text{Tr}(\bm{M}) & 0 \\ 0 & \text{Tr}(\bm{M})
\end{pmatrix} = \bm{Q} + \frac{\text{Tr}(\bm{M})}{2}\bm{I},
\end{equation}
where $\text{Tr}(\bm{M}) = M_{xx} + M_{yy}$, $M_1 = (M_{xx} - M_{yy})/2$, and $M_2 = M_{xy}$. We get the dynamics of the trace by adding together the diagonal components of the ActoMyosin tensor equation \eqref{eq:myosin_equation}:
\begin{equation} \label{eq:TraceDynamics}
\begin{split}
\tau_m(\partial_t \text{Tr}(\bm{M}) + {\bf{v}}\cdot\nabla\text{Tr}(\bm{M})) = 2 & - \text{Tr}(\bm{M}) - \frac{1}{2}\text{Tr}(\bm{M})\text{Tr}(e^{-k_0\bm{\sigma}}) - M_1([e^{-k_0\bm{\sigma}}]_{xx} - [e^{-k_0\bm{\sigma}}]_{yy}) \\ & - 2M_2[e^{-k_0\bm{\sigma}}]_{xy} + D\nabla^2\text{Tr}(\bm{M}),
\end{split}
\end{equation}
where $\bm{\sigma}$ is the total stress. We can get the dynamics for $\bm{Q}$ by taking the traceless part of \eqref{eq:myosin_equation}
\begin{equation} \label{eq:MyosinMinusTrace}
\begin{split}
\tau_m\bigg(\partial_t\bm{M} &+ {\bf{v}}\cdot\nabla\bm{M} + \bm{\omega}\cdot\bm{M} - \bm{M}\cdot\bm{\omega} - \frac{1}{2}[\partial_t \text{Tr}(\bm{M}) + {\bf{v}}\cdot\nabla\text{Tr}(\bm{M})]\bm{I}\bigg)  \\ & = \bm{I} - (\bm{I} + e^{-k_0\bm{\sigma}})\cdot\bm{M} + D\nabla^2\bm{M} - \frac{1}{2}\text{Tr}(\bm{I} - (\bm{I} + e^{-k_0\bm{\sigma}})\cdot\bm{M})\bm{I} -\frac{1}{2}D\nabla^2\text{Tr}(\bm{M})\bm{I},
\end{split}
\end{equation}
where $\bm{\omega} = (\nabla\bm{u} - (\nabla\bm{u})^{\text{T}})/2$ is the anti-symmetric vorticity tensor. Writing this in terms of $\bm{Q}$, we have 
\begin{equation} \label{eq:traceless_myo}
\begin{aligned}
\partial_t\bm{Q} & + {\bf{v}}\cdot\nabla\bm{Q} + \bm{\omega}\cdot\bm{Q} - \bm{Q}\cdot\bm{\omega} \\ & = -(\bm{I} + e^{-k_0\bm{\sigma}})\cdot\bm{Q} - \frac{1}{2}\text{Tr}(\bm{M})e^{-k_0\bm{\sigma}} + \frac{1}{2}\bigg(\text{Tr}(e^{-k_0\bm{\sigma}}\cdot\bm{Q}) + \frac{1}{2}\text{Tr}(\bm{M})\text{Tr}(e^{-k_0\bm{\sigma}})\bigg) + D\nabla^2\bm{Q}.
\end{aligned} 
\end{equation} 

\subsection{Expanding the matrix exponential in powers of Q}
The matrix exponential is
\begin{equation}
\begin{aligned}
 e^{-k_0\bm{\sigma}} & = \sum_{k = 0}^{\infty}\frac{(-k_0\bm{\sigma})^k}{k!} = \bm{I} - k_0\bm{\sigma} + \frac{k_0^2\bm{\sigma}^2}{2} + \mathcal{O}(\bm{\sigma}^3)
\\ & = \bm{I} - k_0(\bm{\pi} + \beta(\bm{Q} + (1/2)\text{Tr}(\bm{M})\bm{I} - m_0\bm{I})) + \frac{k_0^2}{2}(\bm{\pi} + \beta(\bm{Q} + (1/2)\text{Tr}(\bm{M})\bm{I} - m_0\bm{I}))^2
\\ & = \bm{I} - k_0(\bm{\pi} + \beta(\bm{Q} + \tilde{m}_0\bm{I})) + \frac{k_0^2}{2}(\bm{\pi} + \beta(\bm{Q} +  \tilde{m}_0\bm{I}))^2 \\ & = \bm{I} - k_0(\bm{\pi} + \beta(\bm{Q} + \tilde{m}_0\bm{I})) + \frac{k_0^2}{2}(\bm{\pi}^2 + \beta\bm{\pi}\cdot(\bm{Q} + \tilde{m}_0\bm{I}) + \beta(\bm{Q} + \tilde{m}_0\bm{I})\cdot\bm{\pi} + \beta^2(\bm{Q} + \tilde{m}_0\bm{I})^2),
\end{aligned}
\end{equation} 
where 
\begin{equation} \label{eq:tilde_m0}
\tilde{m}_0 = \frac{1}{2}\text{Tr}(\bm{M}) - m_0. 
\end{equation}
From here, we will use the fact that 
\begin{equation} \label{eq:trace_Q_sq}
\bm{Q}^2 = (M_1^2 + M_2^2)\bm{I} = \frac{1}{2}\text{Tr}(\bm{Q}^2)\bm{I}.
\end{equation}
Rearranging terms slightly, and using equation \eqref{eq:trace_Q_sq}, we have
\begin{equation}
\begin{aligned}
e^{-k_0\bm{\sigma}} & = \bm{I} - k_0(\bm{\pi} + \beta(\bm{Q} + \tilde{m}_0\bm{I})) \\ & + \frac{k_0^2}{2}\bigg[\bm{\pi}^2 + \beta(\bm{\pi}\cdot\bm{Q} + \bm{Q}\cdot\bm{\pi}) + 2\beta\tilde{m}_0\bm{\pi} + \beta^2\bigg(2\tilde{m}_0\bm{Q} + \frac{1}{2}\text{Tr}(\bm{Q}^2)\bm{I} + \tilde{m}_0^2\bm{I}\bigg)\bigg] + \mathcal{O}(\bm{Q}^3)
\end{aligned}
\end{equation}
The quantity $e^{-k_0\bm{\sigma}}\cdot\bm{Q}$ is also useful: 
\begin{equation}
\begin{aligned}
e^{-k_0\bm{\sigma}}\cdot\bm{Q} & = \bm{Q} - k_0(\bm{\pi}\cdot\bm{Q} + \beta(\bm{Q}^2 + \tilde{m}_0\bm{Q})) \\ & + \frac{k_0^2}{2}\bigg[\bm{\pi}^2\cdot\bm{Q} + \beta(\bm{\pi}\cdot\bm{Q}^2 + \bm{Q}\cdot\bm{\pi}\cdot\bm{Q}) + 2\beta\tilde{m}_0\bm{\pi}\cdot{Q} + \beta^2\bigg(2\tilde{m}_0\bm{Q}^2 + \tilde{m}_0^2\bm{Q}\bigg)\bigg] + \mathcal{O}(\bm{Q}^3) \\ & = \bm{Q} - k_0(\bm{\pi}\cdot\bm{Q} + \beta(\frac{1}{2}\text{Tr}(\bm{Q}^2)\bm{I} + \tilde{m}_0\bm{Q})) \\ & + \frac{k_0^2}{2}\bigg[\bm{\pi}^2\cdot\bm{Q} + \beta(\frac{1}{2}\text{Tr}(\bm{Q}^2)\bm{\pi} + \bm{Q}\cdot\bm{\pi}\cdot\bm{Q}) + 2\beta\tilde{m}_0\bm{\pi}\cdot\bm{Q} + \beta^2\bigg(\tilde{m}_0\text{Tr}(\bm{Q}^2)\bm{I} + \tilde{m}_0^2\bm{Q}\bigg)\bigg] + \mathcal{O}(\bm{Q}^3)
\end{aligned}
\end{equation}
We can rearrange the terms on the RHS of equation \eqref{eq:traceless_myo}:
\begin{equation} \label{eq:traceless_myo_reordered}
\begin{aligned}
\partial_t\bm{Q} & + {\bf{v}}\cdot\nabla\bm{Q} + \bm{\omega}\cdot\bm{Q} - \bm{Q}\cdot\bm{\omega} \\ & = - \bm{Q} - \bigg[e^{-k_0\bm{\sigma}}\cdot\bm{Q} + \frac{1}{2}\text{Tr}(\bm{M})e^{-k_0\sigma})\bigg] + \frac{1}{2}\text{Tr}\bigg[e^{-k_0\bm{\sigma}}\cdot\bm{Q} + \frac{1}{2}\text{Tr}(\bm{M})e^{-k_0\sigma})\bigg]\bm{I}.
\end{aligned} 
\end{equation} 
The terms proportional to the identity vanish:
\begin{equation*}
a\bm{I} - \frac{1}{2}\text{Tr}(a\bm{I}) = 0.
\end{equation*}
Using \eqref{eq:tilde_m0} to replace $\text{Tr}(\bm{M})$, and grouping the lowest order terms, which are proportional to $\bm{Q}$, $\bm{\pi}$, $\bm{\pi}\cdot\bm{Q}$, and $\bm{Q}\cdot\bm{\pi}$, we have
\begin{equation}
\begin{aligned}
\partial_t\bm{Q} & + {\bf{v}}\cdot\nabla\bm{Q} + \bm{\omega}\cdot\bm{Q} - \bm{Q}\cdot\bm{\omega} \\ & = \bigg[-2 + \beta k_0\bigg(2\tilde{m}_0 + m_0 -\beta k_0\bigg(\frac{3}{2}\tilde{m}_0^2 + \tilde{m}_0m_0\bigg)\bigg)\bigg]\bm{Q} \\ 
& + \bigg[k_0(\tilde{m}_0 + m_0) - \beta k_0 \bigg(k_0\tilde{m}_0(\tilde{m}_0 + m_0) + \frac{1}{4}k_0\text{Tr}(\bm{Q}^2)\bigg)\bigg]\bigg(\bm{\pi} - \frac{1}{2}\text{Tr}(\bm{\pi})\bigg) \\
& + \bigg[k_0 - \frac{1}{2}\beta k_0^2(3\tilde{m}_0 + m_0)\bigg]\bigg(\bm{\pi}\cdot\bm{Q} - \frac{1}{2}\text{Tr}(\bm{\pi}\cdot\bm{Q})\bigg) + \bigg[\frac{1}{2}\beta k_0^2 (\tilde{m}_0 + m_0)\bigg]\bigg(\bm{Q}\cdot\bm{\pi} - \frac{1}{2}\text{Tr}(\bm{Q}\cdot\bm{\pi})\bigg) + \cdots
\end{aligned}
\end{equation}
Taking $\tilde{m}_0$ to be zero, and setting $m_0 = 1/2$ gives 
\begin{equation}
\begin{aligned}
\partial_t\bm{Q} & + {\bf{v}}\cdot\nabla\bm{Q} + \bm{\omega}\cdot\bm{Q} - \bm{Q}\cdot\bm{\omega} \\ & = \bigg[\frac{\beta k_0}{2} - 2\bigg]\bm{Q} +  \frac{k_0}{2}\bigg[1 - \frac{\beta k_0}{2}\text{Tr}(\bm{Q}^2)\bigg]\bigg(\bm{\pi} - \frac{1}{2}\text{Tr}(\bm{\pi})\bigg) \\
& + k_0\bigg[1 - \frac{\beta k_0}{4}\bigg]\bigg(\bm{\pi}\cdot\bm{Q} - \frac{1}{2}\text{Tr}(\bm{\pi}\cdot\bm{Q})\bigg) + \frac{\beta k_0^2}{4}\bigg(\bm{Q}\cdot\bm{\pi} - \frac{1}{2}\text{Tr}(\bm{Q}\cdot\bm{\pi})\bigg) + \cdots
\end{aligned}
\end{equation}

\end{document}